\documentclass[amssymb,aps]{revtex4-2}
\usepackage{graphicx}
\usepackage{enumerate}
\usepackage{amssymb}
\usepackage{amsmath}
\usepackage{amsfonts}
\usepackage{color}
\usepackage{mathrsfs}

\begin{document}
\title{Quantum interference effects in two-photon scattering by a macroscopic lossy sphere}

\author{A. Ciattoni$^1$}
\email{alessandro.ciattoni@spin.cnr.it}
\affiliation{$^1$CNR-SPIN, c/o Dip.to di Scienze Fisiche e Chimiche, Via Vetoio, 67100 Coppito (L'Aquila), Italy}

\date{\today}

\begin{abstract}
We investigate the quantum optical scattering of two-photon wavepackets by a macroscopic lossy sphere by means of macroscopic quantum electrodynamics in the form of modified Langevin noise formalism. The two ingoing photons with arbitrary frequency-polarization spectrum impinge onto the sphere along two different directions and, as consequence of matter losses, their scattering involves the three independent processes where two, one and zero outgoing photons survive. Non-collinearity of ingoing photons causes the existence of two different quantum paths they can follow upon scattering, this producing interference effects in  the detection of the above three processes which is governed by the wavepacket spectral symmetry. By exploiting rotational invariance, we show that different classes of scattering geometries exist such that the coincidence detection of the scattered photons shows perfect constructive or destructive (Hong-Ou-Mandel) interference, both for symmetric and antisymmetric wavepackets. To assess the impact of matter dispersion/losses on quantum interference effects accompanying photons detection, we analyze the scattering of narrow band two-photon wavepackets by high-index dielectric lossy spheres. We show that classical Mie resonance peaks, due to their Fano-like traits, yield very strong constructive and destructive interference effects, occurring when the wavepacket carrier frequency matches the resonance frequency and side Fano dip frequency, respectively. In addition we consider the overall scattering probabilities of two, one and zero photons and we prove that, at the Mie resonance frequencies, they exhibit quantum interference effects which are extremely sensitive to the spectral symmetry of the input wavepacket, thus suggesting an efficient spectral technique assisted by matter losses to identify entanglement.
\end{abstract}
\maketitle

\section{Introduction}
Multi-photon interferometry \cite{Ouuuu1} is among the most relevant research topics in quantum optics, both for its conceptual interest stemming from the intimate connection with photon indistinguishability, and for its far-reaching applications in quantum computation \cite{Nielse} which hinge upon the marked sensitivity to entaglement. The interfering quantum paths are usually provided in literature by lossless optical devices as beam splitters \cite{Prasa1,Ouuuu2,Fearn1,Luiss1}, general passive optical systems \cite{Knoll1} and planar devices \cite{Leyyy3,Savas1,Khanb0,Atama1} which essentially couple their input and output photonic modes without altering the energy balance, i.e. they are characterized by a unitary relation joining the boson operators of output modes to those of the input ones. Lossless beam splitters are able to perform two-photon interference \cite{Hongg1,Fearn2,Ouuuu3,Rarit1,Leger1,Wangg1,Rayme1,Bouch1} which is probably the most paradigmatic quantum interferometric phenomenon, encompassing the limiting effects of photon coalescence, or Hong-Ou-Mandel (HOM) effect, and photon anticoalescence resulting from input two-photon wavepackets whose spectrum is symmetric and antisymmetric, respectively. On the other hand, other lossless devices as Mach-Zehnder interferometers \cite{Larch1,Kimmm1}, plasmonic circuits \cite{Heere1,DiMar1,Fakon1}, silicon chips \cite{Silve1} and q-plates \cite{Nagal1,Dambr1} have been used to produce two-photon interference, all mimicking the archetypal functionality of lossless beam splitters. In addition, lossless devices with more than two input and output ports \cite{Weihs1,Tichy1,Spagn1,Tillm1} have opened the possibility to observe multi-photon interference. When losses can not be neglected, optical devices still couple their input and output photonic modes but radiation-matter energy exchange requires to account for matter modes as well, so that the device is characterized by a unitary input-output relation involving both photon and matter (Langevin) boson operators
\cite{Barne1,Tisch1}, as specifically reported for four-ports \cite{Knoll2} and N-ports \cite{Herna1} devices, multilayer dielectric plates \cite{Grune1} and gratings \cite{Wangg2}. The simplest but fundamental setup of a lossy beam splitter fed by a two-photon wavepacket has been investigated in Refs.\cite{Barne2,Jeffe1} where the authors proved that conditions still exists to observe the HOM effect. In the same paper, the authors also showed that, as a consequence of matter losses, the beam splitter functionality involves the additional processes where one or zero photons survives and that, remarkably, a further quantum interference effect (apparent nonlinear absorption) exists which is able to suppress the one-photon survival process. However, in Ref.\cite{Uppuu1} the authors theoretically argued that the phenomenology of two-photon interference produced by lossy devices is much more rich since the additional freedom provided by losses in designing the device can be harnessed to engineer the interference, making it range from photon coalescence to anticoalescence regardless of the spectral symmetry of the input wavepacket. This is possible since losses literally creates otherwise not existing quantum paths, as experimentally confirmed in various setups \cite{Wolte1,Vestt2,Liiii1,Vetlu1,Wangg3,Hongg2}.

Theoretical predictions concerning two-photon interference rely on the quantum description of the optical device functionality, in turn obtained from a theoretical scheme accounting for the interaction of the quantized electromagnetic field with macroscopic matter. While macroscopic classical electrodynamics is rather simply quantized when only transparent dielectrics are involved \cite{Glaub1,Dalto1}, it is considerably harder to turn it into a quantum theory when optical absorption cannot be neglected, as a consequence of the necessary inclusion in the theory of the matter degrees of freedom \cite{Huttn2,Sutto1,Kheir1,Kheir2,Bhattt,Sutto3}. Langevin noise formalism (LNF) \cite{Grune2,Schee3,Dungg1,Schee4,Philbi}, or macroscopic quantum electrodynamics (MQED), is nowadays the most general theoretical framework for modeling the quantum optical field inside a spatially unbounded absorbing medium, as proved by its many and different applications
\cite{Buhma2,Schee5,Hakam1,Rivera,Hemmer,Kurman,Feistt,DiGiu1,Hayun1,Ciatt1,Ciatt2}. Basically, LNF regards the field as entirely due to dipolar medium sources whose quanta, the electric and magnetic bosonic polaritons, provide the quantum field description. On the other hand, if the absorbing medium is confined within a finite-size object, LNF can still be used by resorting to a singular limiting procedure consisting in taking the permittivity and permeability limits ${\rm Im} ( \varepsilon) \rightarrow 0^+$ and ${\rm Im} ( \mu) \rightarrow 0^+$ pertaining the regions filled by vacuum at the end of the calculations \cite{Hanso1}. Recently, an alternative version of MQED has been introduced, the modified Langevin noise formalism (MLNF) \cite{Naaaa1,Ciatt3}, which avoids the singular limiting procedure when dealing with lossy objects of finite size in vacuum. This is possible since in the MLNF the field displays both the contributions of the dipolar medium sources and of the the scattering modes, so that the quantum portrayal is provided by electric (e), magnetic (m) and scattering (s) bosonic polaritons, the latters being the counterparts of standard photons in the lossy regime. Note that, in the lossless limit, the MLNF coincides with the Glauber's approach to quantum optics of dielectric media of Ref.\cite{Glaub1} since $e$- and $m$-polaritons disappear and $s$ polaritons reduces to standard photons. The MLNF flexibility to deal with the quantized scattering modes has recently been exploited to model the interaction of quantum emitters with dispersive objects \cite{Mianoo} and to develop a general approach to quantum optical scattering by finite lossy objects in vacuum \cite{Ciatt4}.

In this paper we investigate various quantum interference effects accompanying the scattering of two-photon wavepackets by a non-magnetic macroscopic lossy sphere by resorting to the general approach of Ref.\cite{Ciatt4}. It is worth pointing out that the only quantum optical theoretical analysis dealing with a macroscopic sphere is, to the best of our knowledge, the one reported in Ref.\cite{Maure1} where the sphere is assumed to be lossless and dispersionless, as opposed to the present paper where we mainly focus on the role played matter losses on quantum interference effects. After reviewing the general formalism, we consider the scattering by an arbitrary object of two ingoing $s$ polaritons (the photon counterparts in the presence of lossy matter, in MLNF) traveling along different directions and with an arbitrary frequency-polarization wavepacket spectrum. Since matter absorption can destroy an ingoing $s$ polariton with the creation of an $e$ polariton (i.e. a quantized medium dipolar excitation) the outgoing quantum radiation state involves three contributions, respectively containing two $s$, one $s$ and one $e$, and two $e$ outgoing polaritons. The corresponding $ss$, $se$ and $ee$ processes closely parallel those supported by a lossy beam splitter and discussed in Refs.\cite{Barne2,Jeffe1} with the essential difference that, in our case, the directions of the two ingoing $s$ polaritons and those of the detected outgoing $s$ polaritons can be arbitrarly choosen, so that the scatterer object can  effectively be regarded as an optical device with an infinite number of ports. The quantum amplitudes of the three processes turn out to be the superposition of a direct and an exchange term which correspond to the two possible quantum paths connecting the ingoing polaritons to the outgoing ones.  Accordingly the $ss$, $se$ and $ee$ processes display interference effects which are maximal for ingoing wavepacket with symmetric or antisymmetric spectra, so that we here focus on such two relevant cases by analyzing the ensuing probability density of polariton coincidence detection and overall scattering probabilities of two, one and zero polaritons. After specializing the treatment to a macroscopic lossy sphere, we exploit the ensuing full rotational invariance to discuss two classes of scattering geometries such that the polariton coincidence detection exhibits perfect constructive or destructive interference, depending on the wavepacket symmetry (the former effect substantially representing a far-reaching generalization of the famous Hong-Ou-Mandel effect in $50/50$ beam splitters). In order to discuss the role played by matter dispersion/losses of quantum interference effects, we focus on subwavelength lossy spheres with high-index dielectric behavior and, using narrow band ingoing wavepackets, we analyze the spectral features produced by such spheres in the polaritons detection. As a main general result, we prove that the classic Mie resonances of the sphere support very efficient quantum interference mechanism accompanying any measurement concerning the scattered radiation. In particular, in the probability density of polariton coincidence detection, we find that the well-known Fano-like spectral profile of the Mie resonance peaks \cite{Tzaro1,Marmo1} is responsible for very marked constructive or destructive interference respectively occurring when the wavepacket central frequency nearly matches the resonance frequency or the side Fano dip frequency. Besides, in the overall scattering probabilities of two, one and zero $s$-polaritons, we find that the wavepacket symmetry plays a dominant role at isolated Mie resonances where the scattering probabilities for symmentric and antisimmetric wavepackets can differ by several order of magnitude. This suggests that classical resonances can provide efficient alternative ways to detect polariton entanglement.

\section{Main Formalism}
We start by reviewing the MLNF detailed in Ref.\cite{Ciatt3} and its application to quantum optical scattering discussed in Ref.\cite{Ciatt4}, suitably specialized here to a non-magnetic object in vacuum, occupying a finite volume $V$ and whose optical response is characterized in the frequency domain ($e^{ - i\omega t}$) by the dielectric permittivity $\varepsilon _\omega ^V \left( {\bf{s}} \right)$ which is a holomorphic function of $\omega$ for ${\mathop{\rm Im}\nolimits} \left( \omega  \right) > 0$. We denote with ${\varepsilon _\omega  \left( {\bf{s}} \right)}$ the overall permittivity coinciding with $\varepsilon _\omega ^V \left( {\bf{s}} \right)$ within the object (${\bf{s}} \in V$) and being equal to $1$ in vacuum (${\bf{s}} \notin V$), as depicted in Fig.1(a).

\subsection{Modified Langevin noise formalism}
As detailed in Ref.\cite{Ciatt3}, the basic tools MLNF borrows from classical electrodynamics are the modal dyadic ${\cal F}_\omega  \left( {\left. {\bf s} \right|{\bf{n}}} \right)$ and the dyadic Green's function ${\cal G}_\omega  \left( {\left. {\bf s} \right|{\bf{r}}} \right)$ which are defined by the boundary-value problems
\begin{align} \label{GS}
\left[ {\left( {\nabla _{\bf{s}}  \times \nabla _{\bf{s}}  \times } \right) - k_\omega ^2 \varepsilon _\omega  \left( {\bf{s}} \right)} \right]{\cal F}_\omega  \left( {\left. {\bf{s}} \right|{\bf{n}}} \right) &= 0,   & 
{\cal F}_\omega  \left( {\left. {s{\bf{N}}} \right|{\bf{n}}} \right) & \mathop  \approx \limits_{s \to  + \infty }  {e^{i\left( {k_\omega  s} \right){\bf{N}} \cdot {\bf{n}}} {\cal I}_{\bf{n}}  + \frac{{e^{ik_\omega  s} }}{s}{\cal S}_\omega  \left( {\left. {\bf{N}} \right|{\bf{n}}} \right)}, \nonumber \\
\left[ {\left( {\nabla _{\bf{s}}  \times \nabla _{\bf{s}}  \times } \right) - k_\omega ^2 \varepsilon _\omega  \left( {\bf{s}} \right)} \right]{\cal G}_\omega  \left( {\left. {\bf{s}} \right|{\bf{r}}} \right) &= \delta \left( {{\bf{s}} - {\bf{r}}} \right){\cal I}, &
{\cal G}_\omega  \left( {\left. {s{\bf{N}}} \right|{\bf{r}}} \right) &\mathop  \approx \limits_{s \to  + \infty } \frac{{e^{ik_\omega  s} }}{s}{\cal W}_\omega  \left( {\left. {\bf{N}} \right|{\bf{r}}} \right), 
\end{align}
where $k_\omega   = \omega /c$ is the vacuum wavenumber, $\bf N$ and $\bf n$ are unit vectors, ${\cal I}$ is the identity dyadic, ${\cal I}_{\bf{n}}  = {\cal I} - {\bf{nn}}$ is the dyadic projector onto the plane orthogonal to $\bf n$ and the symbol $\mathop  \approx \limits_{s \to + \infty }$ stands for the leading-order term of the asymptotic expansion for $s \to  + \infty$. We hereafter denote with $do_{\bf{n}}  = \sin \theta _{\bf{n}} d\theta _{\bf{n}} d\varphi _{\bf{n}}$  the differential of the solid angle around the unit vector ${\bf{n}} = \sin \theta _{\bf{n}} \left( {\cos \varphi _{\bf{n}} {\bf{u}}_x  + \sin \varphi _{\bf{n}} {\bf{u}}_y } \right) + \cos \theta _{\bf{n}} {\bf{u}}_z$. The modal dyadic satisfies the orthogonality relation ${\cal F}_\omega  \left( {\left. {\bf s} \right|{\bf{n}}} \right) \cdot {\bf{n}} = 0$ and the dyadic ${\cal S}_\omega  \left( {{\bf{N}}\left| {\bf{n}} \right.} \right)$, appearing in its asymptotic behavior, is the classical scattering dyadic \cite{Krist1} satisfying the orthogonality relations ${\bf{N}} \cdot {\cal S}_\omega  \left( {{\bf{N}}\left| {\bf{n}} \right.} \right) = 0$ and ${\cal S}_\omega  \left( {{\bf{N}}\left| {\bf{n}} \right.} \right) \cdot {\bf{n}} = 0$. The dyadic ${\cal W}_\omega  \left( {{\bf{N}}\left| {\bf{r}} \right.} \right)$ is the asympotic amplitude of the dyadic Green's function satisfying the orthogonal relation ${\bf{N}} \cdot {\cal W}_\omega  \left( {{\bf{N}}\left| {\bf{r}} \right.} \right) = 0$ and it is related to the modal dyadic by the remarkable relation ${\cal W}_\omega  \left( {{\bf{N}}\left| {\bf{r}} \right.} \right) = \frac{1}{{4\pi }}{\cal F}_\omega ^T \left( {\left. {\bf{r}} \right| - {\bf{N}}} \right)$ which has been proved in Ref.\cite{Ciatt3} and used there to analytically substantiate the MLNF. The modal dyadic and the dyadic Green's function provide the description of any classical scattering and radiation process in the presence of the object since, if ${\bf{E}}_\omega ^{\left( {in} \right)} \left( {\bf{s}} \right) = \int {do_{\bf{n}} } e^{i\left( {k_\omega  {\bf{n}}} \right) \cdot {\bf{s}}} {\bf{\tilde E}}_\omega ^{\left( {in} \right)} \left( {\bf{n}} \right)$ (with ${\bf{n}} \cdot {\bf{\tilde E}}_\omega ^{\left( {in} \right)} \left( {\bf{n}} \right) = 0$) is any free field impinging onto the object and ${\bf{J}}_\omega  \left( {\bf{s}} \right)$ is the current density of any radiation source, the overall field throughout space is
\begin{equation} 
{\bf{E}}_\omega  \left( {\bf{s}} \right) = \int {do_{\bf{n}} } {\cal F}_\omega  \left( {\left. {\bf{s}} \right|{\bf{n}}} \right) \cdot {\bf{\tilde E}}_\omega ^{\left( {in} \right)} \left( {\bf{n}} \right) + i\omega \mu _0 \int {d^3 {\bf{r}}} \;{\cal G}_\omega  \left( {\left. {\bf{s}} \right|{\bf{r}}} \right) \cdot {\bf{J}}_\omega  \left( {\bf{r}} \right),
\end{equation}
i.e. the superposition of the scattered (first contribution) and radiated (second contribution) fields. From the right column of Eqs.(\ref{GS}), the asymptotic behavior of such field is ${\bf{E}}_\omega  \left( {s{\bf{N}}} \right)\mathop  \approx \limits_{s \to  + \infty } {\bf{E}}_\omega ^{\left( {in} \right)} \left( {s{\bf{N}}} \right) + \frac{{e^{ik_\omega  s} }}{s}\left[ {{\bf{f}}_\omega ^{sca} \left( {\bf{N}} \right) + {\bf{f}}_\omega ^{rad} \left( {\bf{N}} \right)} \right]$ where 
\begin{align}
{\bf{f}}_\omega ^{sca} \left( {\bf{N}} \right) &= \int {do_{\bf{n}} } {\cal S}_\omega  \left( {\left. {\bf{N}} \right|{\bf{n}}} \right) \cdot {\bf{\tilde E}}_\omega ^{\left( {in} \right)} \left( {\bf{n}} \right), &
{\bf{f}}_\omega ^{rad} \left( {\bf{N}} \right) &= i\omega \mu _0 \int {d^3 {\bf{r}}} \;{\cal W}_\omega  \left( {\left. {\bf{N}} \right|{\bf{r}}} \right) \cdot {\bf{J}}_\omega  \left( {\bf{r}} \right)
\end{align}
are its far field scattering and radiation amplitudes.

We hereafter denote with ${\bf{u}}_\upsilon$  the three ($\upsilon = 1,2,3$) cartesian unit vectors spanning the space ($\sum\nolimits_\upsilon  {{\bf{u}}_\upsilon  {\bf{u}}_\upsilon  }  = {\cal I}$) and with ${\bf{e}}_{{\bf{n}}\lambda }$ two ($\lambda = 1,2$) orthonormal real unit vectors orthogonal to $\bf n$ (${\bf{e}}_{{\bf{n}}\lambda }  \cdot {\bf{e}}_{{\bf{n}}\lambda '}  = \delta _{\lambda \lambda '}$, ${\bf{e}}_{{\bf{n}}\lambda }  \cdot {\bf{n}} = 0$) and hence spanning the plane orthogonal to $\bf n$ ($\sum\nolimits_\lambda  {{\bf{e}}_{{\bf{n}}\lambda } {\bf{e}}_{{\bf{n}}\lambda } }  = {\cal I}_{\bf{n}}$). Such unit vectors enable the introduction of the classical fields
\begin{align} \label{FsFe}
{\bf{F}}_s \left( {{\bf s}\left| {\omega {\bf{n}}\lambda } \right.} \right) &= \sqrt {\frac{{\hbar k_\omega ^3 }}{{16 \pi^3 \varepsilon _0 }}} {\cal F}_\omega  \left( {\left. {\bf s} \right|{\bf{n}}} \right) \cdot {\bf{e}}_{{\bf{n}}\lambda }, &
 {\bf{F}}_e \left( {{\bf s}\left| {\omega {\bf{r}}\upsilon } \right.} \right) &= i\sqrt {\frac{{\hbar k_\omega ^4 }}{{\pi \varepsilon _0 }}{\mathop{\rm Im}\nolimits} \left[ {\varepsilon _\omega  \left( {\bf{r}} \right)} \right]} {\cal G}_\omega  \left( {\left. {\bf s} \right|{\bf{r}}} \right) \cdot {\bf{u}}_\upsilon 
\end{align}
in such a way that, apart from the numerical coefficients introduced for later quantization purposes, ${\bf{F}}_s \left( {{\bf{s}}\left| {\omega {\bf{n}}\lambda } \right.} \right)$ is the field excitation at the point ${\bf s}$ resulting from the scattering by the object of an incident vacuum plane wave of frequency $\omega$, direction $\bf n$ and polarization ${\bf{e}}_{{\bf{n}}\lambda }$, whereas ${\bf{F}}_e \left( {{\bf{s}}\left| {\omega {\bf{r}}\upsilon } \right.} \right)$ is the field excitation at the point ${\bf s}$ produced by an electric dipole of frequency $\omega$, located at the point $\bf r$ within the object (since $
{\mathop{\rm Im}\nolimits} \left[ {\varepsilon _\omega  \left( {\bf{r}} \right)} \right] = 0$ in vacuum) and direction ${\bf{u}}_\upsilon$. The MLNF Hamiltonian operator and Heisenberg-picture electric field operator are \cite{Ciatt3} 
\begin{eqnarray} \label{MLFN_HE}
\hat H &=& \int {d\omega } \hbar \omega \left[ {\int {do_{\bf{n}} } \sum\limits_\lambda  {\hat a_s^\dag  \left( {\omega {\bf{n}}\lambda } \right)\hat a_s \left( {\omega {\bf{n}}\lambda } \right)}  + \int {d^3 {\bf{r}}} \sum\limits_\upsilon  {\hat a_e^\dag  \left( {\omega {\bf{r}}\upsilon } \right)\hat a_e \left( {\omega {\bf{r}}\upsilon } \right)} } \right], \nonumber \\ 
{\bf{\hat E}}\left( {{\bf{s}},t} \right) &=& \int {d\omega } e^{ - i\omega t} \left[ {\int {do_{\bf{n}} } \sum\limits_\lambda  {{\bf{F}}_s \left( {{\bf{s}}\left| {\omega {\bf{n}}\lambda } \right.} \right)\hat a_s \left( {\omega {\bf{n}}\lambda } \right)}  + \int {d^3 {\bf{r}}} \sum\limits_\upsilon  {} {\bf{F}}_e \left( {{\bf{s}}\left| {\omega {\bf{r}}\upsilon } \right.} \right)\hat a_e \left( {\omega {\bf{r}}\upsilon } \right)} \right] + {\rm H.c.} 
\end{eqnarray}
where $\int {d\omega }  = \int_0^{ + \infty } {d\omega }$ denotes integration over positive frequencies whereas  $\hat a_s \left( {\omega {\bf{n}}\lambda } \right)$ and $\hat a_e \left( {\omega {\bf{r}}\upsilon } \right)$ are operators satisfying the bosonic commutation relations
\begin{eqnarray} \label{MLFN_bos}
\left[ {\hat a_s \left( {\omega {\bf{n}}\lambda } \right),\hat a_s^\dag  \left( {\omega '{\bf{n}}'\lambda '} \right)} \right] &=& \delta \left( {\omega  - \omega '} \right)\delta \left( {o_{\bf{n}}  - o_{{\bf{n}}'} } \right)\delta _{\lambda \lambda '}, \nonumber  \\ 
\left[ {\hat a_e \left( {\omega {\bf{r}}\upsilon } \right),\hat a_e^\dag  \left( {\omega '{\bf{r}}'\upsilon '} \right)} \right] &=& \delta \left( {\omega  - \omega '} \right)\delta \left( {{\bf{r}} - {\bf{r}}'} \right)\delta _{\upsilon \upsilon '},
\end{eqnarray}
where $\delta \left( {o_{\bf{n}}  - o_{{\bf{n}}'} } \right) = \delta \left( {\cos \theta _{\bf{n}}  - \cos \theta _{{\bf{n}}'} } \right)\delta \left( {\varphi _{\bf{n}}  - \varphi _{{\bf{n}}'} } \right)$ is the angular delta function, together with the vanishing of all the other possible commutation relations. In other words, MLNF describes the quantum optical field as an assembly of noninteracting bosons of two kinds, the $s$ and $e$ polaritons, which are the quanta associated to the field excitations  ${\bf{F}}_s \left( {{\bf{s}}\left| {\omega {\bf{n}}\lambda } \right.} \right)$ and ${\bf{F}}_e \left( {{\bf{s}}\left| {\omega {\bf{r}}\upsilon } \right.} \right)$, respectively, arising out of the light-matter macroscopic coupling. Evidently, $s$ polaritons can be regarded as the counterparts of photons in the lossy regime whereas $e$ polaritons can be viewed as quanta of the medium dipolar sources.

\subsection{Quantum optical scattering approach}
The approach to quantum optical scattering by lossy objects based on the MLNF and introduced in Ref.\cite{Ciatt4} additionally resorts to the transmission ${\cal T}_{\omega ss} \left( {\left. {\bf{N}} \right|{\bf{n}}} \right)$ \cite{Saxon1}, emission ${\cal E}_{\omega se} \left( {\left. {\bf{N}} \right|{\bf{r}}} \right)$, absorption ${\cal A}_{\omega es} \left( {\left. {\bf{R}} \right|{\bf{n}}} \right)$ and energy redistribution ${\cal Q}_{\omega ee} \left( {\left. {\bf{R}} \right|{\bf{r}}} \right)$ dyadics given by
\begin{align} \label{QOS_ClasDyad}
{\cal T}_{\omega ss} \left( {\left. {\bf{N}} \right|{\bf{n}}} \right) &= \delta \left( {o_{\bf{N}}  - o_{\bf{n}} } \right){\cal I}_{\bf{n}}  + \frac{{ik_\omega  }}{{2\pi }}{\cal S}_\omega  \left( {\left. {\bf{N}} \right|{\bf{n}}} \right),  &
{\cal E}_{\omega se} \left( {\left. {\bf{N}} \right|{\bf{r}}} \right) &=  - \sqrt {4k_\omega ^3 {\mathop{\rm Im}\nolimits} \left[ {\varepsilon _\omega  \left( {\bf{r}} \right)} \right]} {\cal W}_\omega  \left( {\left. {\bf{N}} \right|{\bf{r}}} \right),  \nonumber \\
{\cal A}_{\omega es} \left( {\left. {\bf{R}} \right|{\bf{n}}} \right) &=  - \sqrt {4k_\omega ^3 {\mathop{\rm Im}\nolimits} \left[ {\varepsilon _\omega  \left( {\bf{R}} \right)} \right]} {\cal W}_\omega ^T \left( { - {\bf{n}}\left| {\bf{R}} \right.} \right), & 
{\cal Q}_{\omega ee} \left( {\left. {\bf{R}} \right|{\bf{r}}} \right) &=  - \delta \left( {{\bf{R}} - {\bf{r}}} \right){\cal I} - i\sqrt {4k_\omega ^4 {\mathop{\rm Im}\nolimits} \left[ {\varepsilon _\omega  \left( {\bf{R}} \right)} \right]{\mathop{\rm Im}\nolimits} \left[ {\varepsilon _\omega  \left( {\bf{r}} \right)} \right]} {\cal G}_\omega  \left( {{\bf{R}}\left| {\bf{r}} \right.} \right), \nonumber \\
\end{align}
which are shown in Ref.\cite{Ciatt4} to satisfy the relations
\begin{eqnarray} \label{QOS_EnBal1}
\int {do_{\bf{N}} } {\cal T}_{\omega ss}^{T*} \left( {\left. {\bf{N}} \right|{\bf{n}}} \right) \cdot {\cal T}_{\omega ss} \left( {\left. {\bf{N}} \right|{\bf{n}}'} \right) + \int {d^3 {\bf{R}}} {\cal A}_{\omega es}^{T*} \left( {{\bf{R}}\left| {\bf{n}} \right.} \right) \cdot {\cal A}_{\omega es} \left( {{\bf{R}}\left| {{\bf{n}}'} \right.} \right) &=& \delta \left( {o_{\bf{n}}  - o_{{\bf{n}}'} } \right){\cal I}_{\bf{n}}, \nonumber  \\ 
\int {do_{\bf{N}} } {\cal T}_{\omega ss}^{T*} \left( {\left. {\bf{N}} \right|{\bf{n}}} \right) \cdot {\cal E}_{\omega se} \left( {\left. {\bf{N}} \right|{\bf{r}}} \right) + \int {d^3 {\bf{R}}} {\cal A}_{\omega es}^{T*} \left( {{\bf{R}}\left| {\bf{n}} \right.} \right) \cdot {\cal Q}_{\omega ee} \left( {{\bf{R}}\left| {{\bf{r}}} \right.} \right) &=& 0, \nonumber \\ 
\int {do_{\bf{N}} } {\cal E}_{\omega se}^{T*} \left( {\left. {\bf{N}} \right|{\bf{r}}} \right) \cdot {\cal E}_{\omega se} \left( {\left. {\bf{N}} \right|{\bf{r}}'} \right) + \int {d^3 {\bf{R}}} {\cal Q}_{\omega ee}^{T*} \left( {{\bf{R}}\left| {\bf{r}} \right.} \right) \cdot {\cal Q}_{\omega ee} \left( {{\bf{R}}\left| {{\bf{r}}'} \right.} \right) &=& \delta \left( {{\bf{r}} - {\bf{r}}'} \right){\cal I},
\end{eqnarray}
and
\begin{eqnarray} \label{QOS_EnBal2}
\int {do_{\bf{n}} } {\cal T}_{\omega ss} \left( {\left. {\bf{N}} \right|{\bf{n}}} \right) \cdot {\cal T}_{\omega ss}^{T*} \left( {\left. {{\bf{N}}'} \right|{\bf{n}}} \right) + \int {d^3 {\bf{r}}}  {\cal E}_{\omega se} \left( {\left. {\bf{N}} \right|{\bf{r}}} \right) \cdot {\cal E}_{\omega se}^{T*} \left( {\left. {{\bf{N}}'} \right|{\bf{r}}} \right) &=& \delta \left( {o_{\bf{N}}  - o_{{\bf{N}}'} } \right){\cal I}_{\bf{N}}, \nonumber  \\ 
\int {do_{\bf{n}} } {\cal T}_{\omega ss} \left( {\left. {\bf{N}} \right|{\bf{n}}} \right) \cdot {\cal A}_{\omega es}^{T*} \left( {{\bf{R}}\left| {\bf{n}} \right.} \right) + \int {d^3 {\bf{r}}} {\cal E}_{\omega se} \left( {\left. {\bf{N}} \right|{\bf{r}}} \right) \cdot {\cal Q}_{\omega ee}^{T*} \left( {{\bf{R}}\left| {\bf{r}} \right.} \right) &=& 0, \nonumber  \\  
\int {do_{\bf{n}} } {\cal A}_{\omega es} \left( {\left. {\bf{R}} \right|{\bf{n}}} \right) \cdot {\cal A}_{\omega es}^{T*} \left( {{\bf{R}}'\left| {\bf{n}} \right.} \right) + \int {d^3 {\bf{r}}} {\cal Q}_{\omega ee} \left( {\left. {\bf{R}} \right|{\bf{r}}} \right) \cdot {\cal Q}_{\omega ee}^{T*} \left( {{\bf{R}}'\left| {\bf{r}} \right.} \right) &=& \delta \left( {{\bf{R}} - {\bf{R}}'} \right){\cal I}. 
\end{eqnarray}
Equations (\ref{QOS_EnBal1}) and (\ref{QOS_EnBal2}) physically ensure, in full generality, energy conservation of the field-matter system upon classical scattering, encompassing the situations where the scatterer hosts sources able to radiate and absorb energy. In the quantum optical scattering approach, field-matter descriptions in the far past  ($t \rightarrow -\infty$) and far future  ($t \rightarrow +\infty$)  are respectively afforded by the ingoing polariton operators $\hat a_s \left( {\omega {\bf{n}}\lambda } \right)$, $\hat a_e \left( {\omega {\bf{r}}\upsilon } \right)$  (lowercase letters) and  outgoing polariton operators $\hat A_s \left( {\omega {\bf{N}}\Lambda } \right),\hat A_e \left( {\omega {\bf{R}}\Upsilon } \right)$ (capital letters), the latters being linked to the former ones by the input-output transformation
\begin{eqnarray} \label{QOS-AQ}
\hat A_s \left( {\omega {\bf{N}}\Lambda } \right) &=& \int {do_{\bf{n}} } \sum\limits_\lambda  {\left[ {{\bf{e}}_{{\bf{N}}\Lambda }  \cdot {\cal T}_{\omega ss} \left( {\left. {\bf{N}} \right|{\bf{n}}} \right) \cdot {\bf{e}}_{{\bf{n}}\lambda } } \right]} \hat a_s \left( {\omega {\bf{n}}\lambda } \right) + \int {d^3 {\bf{r}}} \sum\limits_\upsilon  {\left[ {{\bf{e}}_{{\bf{N}}\Lambda }  \cdot {\cal E}_{\omega se} \left( {\left. {\bf{N}} \right|{\bf{r}}} \right) \cdot {\bf{u}}_\upsilon  } \right]} \hat a_e \left( {\omega {\bf{r}}\upsilon } \right), \nonumber \\ 
\hat A_e \left( {\omega {\bf{R}}\Upsilon } \right) &=& \int {do_{\bf{n}} } \sum\limits_\lambda  {\left[ {{\bf{u}}_\Upsilon   \cdot {\cal A}_{\omega es} \left( {{\bf{R}}\left| {\bf{n}} \right.} \right) \cdot {\bf{e}}_{{\bf{n}}\lambda } } \right]} \hat a_s \left( {\omega {\bf{n}}\lambda } \right) + \int {d^3 {\bf{r}}} \sum\limits_\upsilon  {\left[ {{\bf{u}}_\Upsilon   \cdot {\cal Q}_{\omega ee} \left( {{\bf{R}}\left| {\bf{r}} \right.} \right) \cdot {\bf{u}}_\upsilon  } \right]} \hat a_e \left( {\omega {\bf{r}}\upsilon } \right). 
\end{eqnarray}
Note that the linear mixture of the operators $\hat a_s$ and $\hat a_e$ in the RHSs of Eqs.(\ref{QOS-AQ}) is produced by the matrix elements of the classical dyadics of Eqs.(\ref{QOS_ClasDyad}). The crucial point is that Eqs.(\ref{QOS_EnBal1}) and (\ref{QOS_EnBal2}) entail that the input-output trasformation in Eqs.(\ref{QOS-AQ}) is invertible, i.e.
\begin{eqnarray} \label{QOS-aq}
\hat a_s \left( {\omega {\bf{n}}\lambda } \right) &=& \int {do_{\bf{N}} } \sum\limits_\Lambda  {\left[ {{\bf{e}}_{{\bf{n}}\lambda }  \cdot {\cal T}_{\omega ss}^{T*} \left( {\left. {\bf{N}} \right|{\bf{n}}} \right) \cdot {\bf{e}}_{{\bf{N}}\Lambda } } \right]\hat A_s \left( {\omega {\bf{N}}\Lambda } \right)}  + \int {d^3 {\bf{R}}} \sum\limits_\Upsilon  {\left[ {{\bf{e}}_{{\bf{n}}\lambda }  \cdot {\cal A}_{\omega es}^{T*} \left( {{\bf{R}}\left| {\bf{n}} \right.} \right) \cdot {\bf{u}}_\Upsilon  } \right]} \hat A_e \left( {\omega {\bf{R}}\Upsilon } \right), \nonumber \\
\hat a_e \left( {\omega {\bf{r}}\upsilon } \right) &=& \int {do_{\bf{N}} } \sum\limits_\Lambda  {\left[ {{\bf{u}}_\upsilon   \cdot {\cal E}_{\omega se}^{T*} \left( {\left. {\bf{N}} \right|{\bf{r}}} \right) \cdot {\bf{e}}_{{\bf{N}}\Lambda } } \right]\hat A_s \left( {\omega {\bf{N}}\Lambda } \right)}  + \int {d^3 {\bf{R}}} \sum\limits_\Upsilon  {\left[ {{\bf{u}}_\upsilon   \cdot {\cal Q}_{\omega ee}^{T*} \left( {{\bf{R}}\left| {\bf{r}} \right.} \right) \cdot {\bf{u}}_\Upsilon  } \right]\hat A_e \left( {\omega {\bf{R}}\Upsilon } \right)}  
\end{eqnarray}
and that bosonic commutation relations are conserved, i.e.
\begin{eqnarray}
&& \left[ {\hat A_s \left( {\omega {\bf{N}}\Lambda } \right),\hat A_s^\dag  \left( {\omega '{\bf{N}}'\Lambda '} \right)} \right] = \delta \left( {\omega  - \omega '} \right)\delta \left( {o_{\bf{N}}  - o_{{\bf{N}}'} } \right)\delta _{\Lambda \Lambda '}, \nonumber  \\ 
&& \left[ {\hat A_e \left( {\omega {\bf{R}}\Upsilon } \right),\hat A_e^\dag  \left( {\omega '{\bf{R}}'\Upsilon '} \right)} \right] = \delta \left( {\omega  - \omega '} \right)\delta \left( {{\bf{R}} - {\bf{R}}'} \right)\delta _{\Upsilon \Upsilon '},
\end{eqnarray}
toghether with the vanishing of all the other possible commutation relation. In other words, as a consequence of energy conservation upon classical scattering, the input-output transformation of Eqs.(\ref{QOS-AQ}) is unitary and the operators $(\hat a_s, \hat a_e)$ and $(\hat A_s,\hat A_e)$ provide two different representations of the field-matter Fock space. It is worth pointing out that, as detailed in Ref.\cite{Ciatt4}, the far field asymptotics of the electric field operator ${\bf{\hat E}} \left( {{\bf{s}},t} \right)$ (second of Eqs.(\ref{MLFN_HE})) in the far past and far future are respectively given by
\begin{eqnarray}
 {\bf{\hat E}}\left( {\xi {\bf{n}},t} \right) &\mathop  \approx \limits_{\scriptstyle \xi  \to  - \infty  \hfill \atop 
  \scriptstyle t \to  - \infty  \hfill} & \int\limits_0^{ + \infty } {d\omega } \sqrt {\frac{{\hbar k_\omega  }}{{4\pi \varepsilon _0 }}} \frac{{e^{ik_\omega  \left( {\xi  - ct} \right)} }}{{i\xi }}\sum\limits_\lambda  {{\bf{e}}_{{\bf{n}}\lambda } \hat a_s \left( {\omega {\bf{n}}\lambda } \right)}  + {\rm H.c.}, \\ 
 {\bf{\hat E}}\left( {\xi {\bf{N}},t} \right) &\mathop  \approx \limits_{\scriptstyle \xi  \to  + \infty  \hfill \atop 
  \scriptstyle t \to  + \infty  \hfill} & \int\limits_0^{ + \infty } {d\omega } \sqrt {\frac{{\hbar k_\omega  }}{{4\pi \varepsilon _0 }}} \frac{{e^{ik_\omega  \left( {\xi  - ct} \right)} }}{{i\xi }}\sum\limits_\Lambda  {{\bf{e}}_{{\bf{N}}\Lambda } \hat A_s \left( {\omega {\bf{N}}\Lambda } \right)}  + {\rm H.c.}, 
\end{eqnarray}
and their respective identical dependencies on $\hat a_s$ and $\hat A_s$ manifestly proves the above statement that the ingoing operators $(\hat a_s, \hat a_e)$ and the outgoing operators $(\hat A_s,\hat A_e)$ respectively provide the input and output field-matter descriptions in quantum optical scattering. Accordingly, using hereafter the Heisenberg picture, the general field-matter input state is of the form $\left| \Psi  \right\rangle  = F^{\left( {in} \right)} \left( {\hat a_s^\dag  ,\hat a_e^\dag  } \right)\left| 0 \right\rangle$, where ${\left| 0 \right\rangle }$ is the normalized polariton vacuum state and $F^{(in)}$ is a functional accounting for the ingoing $s$ polaritons coming from the far field and the ingoing $e$ polaritons stored in the object volume at $t = -\infty$. Since in the Heisenberg picture the state does not evolve, the output state is obtained by expressing $\left| \Psi  \right\rangle$ in terms of the outgoing polariton creation operators, i.e. by substituting the Hermitian conjugates of Eqs.(\ref{QOS-aq}) into the above expression of $ \left| \Psi  \right\rangle$, thus getting $\left| \Psi  \right\rangle  = F^{\left( {out} \right)} \left( {A_s^\dag  ,A_e^\dag  } \right)\left| 0 \right\rangle$, where the obtained functional $F^{\left( {out} \right)}$ describe the outgoing $s$ polaritons traveling toward the far field and the outgoing $e$ polaritons stored in the object volume at $t = +\infty$. Note that, since the scatterer object has finite size, $s$ polaritons in the far field coincide with standard photons of vacuum quantum electrodynamics; however we hereafter use the nomenclature of ingoing and outgoing $s$ polaritons (instead of photons) as a reminder of the underlying dispersive nature of the scatterer object.

\begin{figure}
\centering
\includegraphics[width = 1\linewidth]{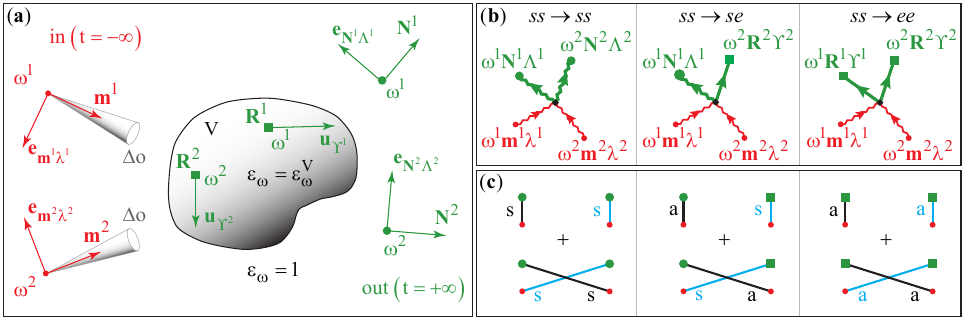}
\caption{(a) Geometry of two directional $ s$ polaritons scattering by a lossy object $V$ of finite size. At $t = -\infty$ (red) the radiation-matter quantum state contains only the pairs $ss$ of ingoing $s$-polaritons (circles) whose directions lies within the two non overlapping cones of solid angles $\Delta o \ll 1$ around the unit vectors ${\bf m}^1$ and ${\bf m}^2$. At $t = +\infty$ (green) the state contains the three pairs $ss$, $se$ and $ee$ of outgoing $s$ (circles) and $e$ (squares) polaritons. (b) Diagramatic representation of the three processes involved in the scattering of the two ingoing $s$ polaritons. (c) Graphical representation of the two possible quantum paths connecting the ingoing $s$ polaritons to the outgoing ones in the three processes.}
\label{Fig1}
\end{figure}

\section{Scattering of two directional $s$ polaritons}
We here exploit the framework reviewed in the last section to analyze the quantum scattering of two non-collinear ingoing $s$ polaritons by an electromagnetically inert lossy object (i.e. not hosting ingoing $e$ polaritons). Besides, we specialize the general analysis to the case of a homogeneous lossy sphere.

\subsection{Input and output radiation-matter descriptions}
In the far past $t = -\infty$, the radiation-matter description is provided by the operators $\hat a_s$ and $\hat a_e$ so that, in order to put no restrictions on the frequency-polarization correlations between the two $s$ polaritons, we assume that the Heisenberg state of the ingoing wavepacket is
\begin{equation} \label{Gen_State}
\left| {\Psi } \right\rangle   = \int {d\omega ^1 d\omega ^2 } \int {do_{{\bf{n}}^1 } } do_{{\bf{n}}^2 } \sum\limits_{\lambda ^1 \lambda ^2 } {\phi _{\lambda ^1 \lambda ^2 } \left( {\omega ^1 ,\omega ^2 } \right)\Theta _{{\bf{m}}^1 } \left( {{\bf{n}}^1 } \right)\Theta _{{\bf{m}}^2 } \left( {{\bf{n}}^2 } \right)\hat a_s^\dag  \left( {\omega ^1 {\bf{n}}^1 \lambda ^1 } \right)\hat a_s^\dag  \left( {\omega ^2 {\bf{n}}^2 \lambda ^2 } \right)\left| 0 \right\rangle },
\end{equation}
where ${\bf{m}}^1$ and ${\bf{m}}^2$ are unit vectors with ${\bf{m}}^1 \neq {\bf{m}}^2$, ${\phi _{\lambda ^1 \lambda ^2 } \left( {\omega ^1 ,\omega ^2 } \right)}$ is a frequency-polarization spectrum and the function $\Theta _{\bf{m}} \left( {\bf{n}} \right) = \frac{1}{{\sqrt {\Delta o} }}{\rm H}\left( {\frac{{\Delta \theta _{\bf{n}} }}{2} - \left| {\theta _{\bf{n}}  - \theta _{\bf{m}} } \right|} \right) {\rm H} \left( {\frac{{\Delta \varphi _{\bf{n}} }}{2} - \left| {\varphi _{\bf{n}}  - \varphi _{\bf{m}} } \right|} \right)$, where $\rm H$ is the Heaviside step function, selects the unit vectors $\bf n$ within a fixed and very small solid angle $\Delta o = \Delta \theta _{\bf{n}} \Delta \varphi _{\bf{n}} \sin \theta _{\bf{n}}  \ll 1$ around the unit vector $\bf m$ and it is such that $\int {do_{\bf{n}} \left[ {\Theta _{\bf{m}} \left( {\bf{n}} \right)} \right]^2 }  = 1$. Accordingly, with reference to Fig.1(a), the state in Eq.(\ref{Gen_State}) is the superposition of all the basis states with two $s$ polaritons whose directions ${\bf n}^1$ and ${\bf n}^2$ lie within the two thin cones (grey) around the directions ${\bf m}^1$ and ${\bf m}^2$ (red). We hereafter assume that such two cones do not overlap, i.e. $\Theta _{{\bf{m}}^1 } \left( {\bf{n}} \right)\Theta _{{\bf{m}}^2 } \left( {\bf{n}} \right) = 0$, so that the state normalization $\left\langle {\Psi }  \mathrel{\left | {\vphantom {\Psi  \Psi }} \right. \kern-\nulldelimiterspace} {\Psi } \right\rangle  = 1$ is easily seen to be assured by the spectral normalization 
\begin{equation} \label{spect_norm}
\int {d\omega ^1 d\omega ^2 } \sum\limits_{\lambda ^1 \lambda ^2 } {\left| {\phi _{\lambda ^1 \lambda ^2 } \left( {\omega ^1 ,\omega ^2 } \right)} \right|^2 }  = 1.
\end{equation}
The ingoing polariton wavefunction $\Psi _{ss}^{\left( {in} \right)} \left( {\omega ^1 {\bf{n}}^1 \lambda ^1 ,\omega ^2 {\bf{n}}^2 \lambda ^2 } \right) = \frac{1}{{\sqrt 2 }}\left\langle 0 \right|\hat a_s \left( {\omega ^1 {\bf{n}}^1 \lambda ^1 } \right)\hat a_s \left( {\omega ^2 {\bf{n}}^2 \lambda ^2 } \right)\left| {\Psi  } \right\rangle $ is given by 
\begin{equation} \label{ing_wavfun}
 \Psi _{ss}^{\left( {in} \right)} \left( {\omega ^1 {\bf{n}}^1 \lambda ^1 ,\omega ^2 {\bf{n}}^2 \lambda ^2 } \right) = \frac{1}{{\sqrt 2 }}\left[ {\phi _{\lambda ^1 \lambda ^2 } \left( {\omega ^1 ,\omega ^2 } \right)\Theta _{{\bf{m}}^1 } \left( {{\bf{n}}^1 } \right)\Theta _{{\bf{m}}^2 } \left( {{\bf{n}}^2 } \right) + \phi _{\lambda ^2 \lambda ^1 } \left( {\omega ^2 ,\omega ^1 } \right)\Theta _{{\bf{m}}^1 } \left( {{\bf{n}}^2 } \right)\Theta _{{\bf{m}}^2 } \left( {{\bf{n}}^1 } \right)} \right]
\end{equation}
and it is evidently symmetric under the exchange $\omega ^1 {\bf{n}}^1 \lambda ^1  \leftrightarrow \omega ^2 {\bf{n}}^2 \lambda ^2$ of the two $s$ polaritons. It is worth pointing out that the non-collinearity of two thin cones of Fig.1(a) ${\bf{m}}^1  \ne {\bf{m}}^2$ causes the wavefunction $\Psi _{ss}^{\left( {in} \right)}$ to be the sum of a direct and an exchange term whose respective amplitudes  ${\phi _{\lambda ^1 \lambda ^2 } \left( {\omega ^1 ,\omega ^2 } \right)}$ and ${\phi _{\lambda ^2 \lambda ^1 } \left( {\omega ^2 ,\omega ^1 } \right)}$ are generally different since the exchange $\omega ^1 \lambda ^1  \leftrightarrow \omega ^2 \lambda ^2$ does not amount to the physical exchange of the two $s$ polaritons. In other words, the spectrum ${\phi _{\lambda ^1 \lambda ^2 } \left( {\omega ^1 ,\omega ^2 } \right)}$ is not required to be exchange-symmetric since the  $s$ polaritons whose directions lie within the cones of Fig.1(a) are distinguishable. 

In the far future $t = +\infty$, the radiation-matter description is provided by the operators $\hat A_s$ and $\hat A_e$  and it can be obtained by elimanting the two creation operators $\hat a_s^\dag$ in the state $\left| \Psi  \right\rangle$ of Eq.(\ref{Gen_State}) by means of the Hermitian conjugate of the first of Eqs.(\ref{QOS-aq}) which is 
\begin{equation}
\hat a_s^\dag  \left( {\omega {\bf{n}}\lambda } \right) = \int {do_{\bf{N}} } \sum\limits_\Lambda  {\left[ {{\bf{e}}_{{\bf{N}}\Lambda }  \cdot {\cal T}_{\omega ss} \left( {\left. {\bf{N}} \right|{\bf{n}}} \right) \cdot {\bf{e}}_{{\bf{n}}\lambda } } \right]\hat A_s^\dag  \left( {\omega {\bf{N}}\Lambda } \right)}  + \int {d^3 {\bf{R}}} \sum\limits_\Upsilon  {\left[ {{\bf{u}}_\Upsilon   \cdot {\cal A}_{\omega es} \left( {{\bf{R}}\left| {\bf{n}} \right.} \right) \cdot {\bf{e}}_{{\bf{n}}\lambda } } \right]} \hat A_e^\dag  \left( {\omega {\bf{R}}\Upsilon } \right).
\end{equation}
The ensuing expression of the state displays three contributions containing $\hat A_s^\dag \hat A_s^\dag$, $\hat A_s^\dag \hat A_e^\dag$ and $\hat A_e^\dag  \hat A_e^\dag $ respectively corresponding to the three physical processes where both $s$ polaritons are scattered ($ss \rightarrow ss$), one is scattered and the other one decays into an $e$ polariton ($ss \rightarrow se$) and both decay into two $e$ polaritons ($ss \rightarrow ee$), as diagramatically represented in Fig.1(b). The amplitudes of the three processes are the outgoing polariton wavefunctions $ \Psi _{ss}^{\left( {out} \right)} \left( {\omega ^1 {\bf{N}}^1 \Lambda ^1 ,\omega ^2 {\bf{N}}^2 \Lambda ^2 } \right) = \frac{1}{{\sqrt 2 }}\left\langle 0 \right|\hat A_s \left( {\omega ^1 {\bf{N}}^1 \Lambda ^1 } \right)\hat A_s \left( {\omega ^2 {\bf{N}}^2 \Lambda ^2 } \right)\left| \Psi  \right\rangle$, $\Psi _{se}^{\left( {out} \right)} \left( {\omega ^1 {\bf{N}}^1 \Lambda ^1 ,\omega ^2 {\bf{R}}^2 \Upsilon ^2 } \right) = \left\langle 0 \right|\hat A_s \left( {\omega ^1 {\bf{N}}^1 \Lambda ^1 } \right)\hat A_e \left( {\omega ^2 {\bf{R}}^2 \Upsilon ^2 } \right)\left| \Psi  \right\rangle$ and $\Psi _{ee}^{\left( {out} \right)} \left( {\omega ^1 {\bf{R}}^1 \Upsilon ^1 ,\omega ^2 {\bf{R}}^2 \Upsilon ^2 } \right) = \frac{1}{{\sqrt 2 }}\left\langle 0 \right|\hat A_e \left( {\omega ^1 {\bf{R}}^1 \Upsilon ^1 } \right)\hat A_e \left( {\omega ^2 {\bf{R}}^2 \Upsilon ^2 } \right)\left| \Psi  \right\rangle$. In order not to overburden the mathematical treatment, we will resort to the following three simplifying, but not very restrictive, assumptions.
\begin{itemize}
\item We omit the delta function contribution in the transmission dyadic ${{\cal T}_{\omega ss} \left( {\left. {\bf{N}} \right|{\bf{n}}} \right)}$ of the first of Eqs.(\ref{QOS_ClasDyad}), this corresponding to analyzing the only scattered field (i.e., without the incident plane wave field contribution). This assumption is strictly valid for outgoing $s$ polaritons whose directions ${\bf N}^1$ and ${\bf N}^2$ do not lie within the two thin cones of Fig.1(a),  i.e. $\Theta _{{\bf{m}}^1 } \left( {{\bf{N}}^1 } \right) = \Theta _{{\bf{m}}^2 } \left( {{\bf{N}}^1 } \right) = 0$ and $\Theta _{{\bf{m}}^1 } \left( {{\bf{N}}^2 } \right) = \Theta _{{\bf{m}}^2 } \left( {{\bf{N}}^2 } \right) = 0$.
\item We assume that the scattering and absorption dyadics ${{\cal S}_\omega  \left( {\left. {\bf{N}} \right|{\bf{n}}} \right)}$ and ${{\cal A}_{\omega es} \left( {\left. {\bf{R}} \right|{\bf{n}}} \right)}$ do not appreciabily vary when $\bf n$ spans the small solid angle $\Delta o$ in such a way that the approximation $\int {do_{\bf{n}} } \Theta _{\bf{m}} \left( {\bf{n}} \right)f\left( {\bf{n}} \right) \simeq \sqrt {\Delta o} f\left( {\bf{m}} \right)$ can be safely used. 
\item We assume that the frequency bandwidth of the spectrum $\phi _{\lambda ^1 \lambda ^2 } \left( {\omega ^1 ,\omega ^2 } \right)$ is narrow enough  that the frequency dependence of ${{\cal S}_\omega  \left( {\left. {\bf{N}} \right|{\bf{n}}} \right)}$ and ${{\cal A}_{\omega es} \left( {\left. {\bf{R}} \right|{\bf{n}}} \right)}$ can be neglected over the bandwidth so that they can be evaluated at a central bandwidth frequency $\omega$. 
\end{itemize}
Under the above assumptions, after tedious but straightforward algebra, the outgoing wavefunctions are seen to be given by
\begin{eqnarray} \label{Psi_ss_se_ee}
\Psi _{ss}^{\left( {out} \right)} \left( {\omega ^1 {\bf{N}}^1 \Lambda ^1 ,\omega ^2 {\bf{N}}^2 \Lambda ^2 } \right) &=& \frac{{\Delta o}}{{\sqrt 2 }}\sum\limits_{\lambda ^1 \lambda ^2 } {\left[ {\phi _{\lambda ^1 \lambda ^2 } \left( {\omega ^1 ,\omega ^2 } \right)s_{{\bf{m}}^1 \lambda ^1 }^{{\bf{N}}^1 \Lambda ^1 } s_{{\bf{m}}^2 \lambda ^2 }^{{\bf{N}}^2 \Lambda ^2 }  + \phi _{\lambda ^1 \lambda ^2 } \left( {\omega ^2 ,\omega ^1 } \right)s_{{\bf{m}}^1 \lambda ^1 }^{{\bf{N}}^2 \Lambda ^2 } s_{{\bf{m}}^2 \lambda ^2 }^{{\bf{N}}^1 \Lambda ^1 } } \right]}, \nonumber  \\ 
\Psi _{se}^{\left( {out} \right)} \left( {\omega ^1 {\bf{N}}^1 \Lambda ^1 ,\omega ^2 {\bf{R}}^2 \Upsilon ^2 } \right) &=& \Delta o\sum\limits_{\lambda ^1 \lambda ^2 } {\left[ {\phi _{\lambda ^1 \lambda ^2 }  \left( {\omega ^1 ,\omega ^2 } \right)s_{{\bf{m}}^1 \lambda ^1 }^{{\bf{N}}^1 \Lambda ^1 } a_{{\bf{m}}^2 \lambda ^2 }^{{\bf{R}}^2 \Upsilon ^2 }  + \phi _{\lambda ^1 \lambda ^2 }  \left( {\omega ^2 ,\omega ^1 } \right)a_{{\bf{m}}^1 \lambda ^1 }^{{\bf{R}}^2 \Upsilon ^2 } s_{{\bf{m}}^2 \lambda ^2 }^{{\bf{N}}^1 \Lambda ^1 } } \right]}, \nonumber  \\ 
\Psi _{ee}^{\left( {out} \right)} \left( {\omega ^1 {\bf{R}}^1 \Upsilon ^1 ,\omega ^2 {\bf{R}}^2 \Upsilon ^2 } \right) &=& \frac{{\Delta o}}{{\sqrt 2 }}\sum\limits_{\lambda ^1 \lambda ^2 } {\left[ {\phi _{\lambda ^1 \lambda ^2 }  \left( {\omega ^1 ,\omega ^2 } \right)a_{{\bf{m}}^1 \lambda ^1 }^{{\bf{R}}^1 \Upsilon ^1 } a_{{\bf{m}}^2 \lambda ^2 }^{{\bf{R}}^2 \Upsilon ^2 }  + \phi _{\lambda ^1 \lambda ^2 }  \left( {\omega ^2 ,\omega ^1 } \right)a_{{\bf{m}}^1 \lambda ^1 }^{{\bf{R}}^2 \Upsilon ^2 } a_{{\bf{m}}^2 \lambda ^2 }^{{\bf{R}}^1 \Upsilon ^1 } } \right]},
\end{eqnarray}
where 
\begin{align} \label{sa}
s_{{\bf{n}}\lambda }^{{\bf{N}}\Lambda } \left( \omega  \right) &= {\bf{e}}_{{\bf{N}}\Lambda }  \cdot \frac{{ik_\omega  }}{{2\pi }}{\cal S}_\omega  \left( {\left. {\bf{N}} \right|{\bf{n}}} \right) \cdot {\bf{e}}_{{\bf{n}}\lambda }, &  a_{{\bf{n}}\lambda }^{{\bf{R}}\Upsilon } \left( \omega  \right) &= {\bf{u}}_\Upsilon   \cdot {\cal A}_{\omega es} \left( {{\bf{R}}\left| {\bf{n}} \right.} \right) \cdot {\bf{e}}_{{\bf{n}}\lambda }
\end{align}
are the scattering and absorption coefficients of the scatterer object. The wavefunctions $\Psi _{ss}^{\left( {out} \right)}$ and $\Psi _{ee}^{\left( {out} \right)}$ evidently display the exchange symmetries $\omega ^1 {\bf{N}}^1 \Lambda ^1  \leftrightarrow \omega ^2 {\bf{N}}^2 \Lambda ^2$  and $\omega ^1 {\bf{R}}^1 \Upsilon ^1  \leftrightarrow \omega ^2 {\bf{R}}^2 \Upsilon ^2$, respectively. The crucial point here is that each of the three wavefunctions in Eqs.(\ref{Psi_ss_se_ee}) is the sum of a direct and an exchance contribution, respectively proportional to the amplitudes ${\phi _{\lambda ^1 \lambda ^2 } \left( {\omega ^1 ,\omega ^2 } \right)}$ and $\phi _{\lambda ^2 \lambda ^1 } \left( {\omega ^2 ,\omega ^1 } \right)$ (after a $\lambda^1 \lambda^2$ relabelling) appearing in the ingoing wavefunction of Eq.(\ref{ing_wavfun}). Moreover, the subscripts and superscripts of the coefficients $s_{{\bf{n}}\lambda }^{{\bf{N}}\Lambda }$ and $a_{{\bf{n}}\lambda }^{{\bf{R}}\Upsilon }$ in Eqs.(\ref{Psi_ss_se_ee}) reveal that, in each process, the direct and exchange contributions correspond to the two possible quantum paths connecting the ingoing $s$ polaritons ${\omega ^1 {\bf{m}}^1 \lambda ^1 }$ and ${\omega ^2 {\bf{m}}^2 \lambda ^2 }$ to the two outgoing ones involved in the process, as graphically sketched in Fig.1(c). Therefore, we conclude that the detection probability of each of the three processes can display quantum interference effects arising from the interplay between the direct and exchange paths, in turn resulting from the ingoing $s$ polaritons non-collinearity (assigning distinct physical meanings to ${\phi _{\lambda ^1 \lambda ^2 } \left( {\omega ^1 ,\omega ^2 } \right)}$ and ${\phi _{\lambda ^2 \lambda ^1 } \left( {\omega ^2 ,\omega ^1 } \right)}$ in Eq.(\ref{ing_wavfun})) and from electromagnetic propagation (provided by the classical scattering and absorpion dyadics ${{\cal S}_\omega  \left( {\left. {\bf{N}} \right|{\bf{n}}} \right)}$ and ${{\cal A}_{\omega es} \left( {\left. {\bf{R}} \right|{\bf{n}}} \right)}$).

\subsection{Wavepackets with definite spectral symmetry}
From Eqs.(\ref{Psi_ss_se_ee}) it is manifest that, for fixed directions ${\bf m}^1$ and ${\bf m}^2$, the ingoing wavepackets producing the largest interference effects are those whose spectrum is such that $\left| {\phi _{\lambda ^2 \lambda ^1 } \left( {\omega ^2 ,\omega ^1 } \right)} \right| = \left| {\phi _{\lambda ^1 \lambda ^2 } \left( {\omega ^1 ,\omega ^2 } \right)} \right|$ so that we hereafter focus on ingoing wavepackets whose spectrum $\phi _{\lambda ^1 \lambda ^2 }^\sigma  \left( {\omega ^1 ,\omega ^2 } \right)$ is either symmetric ($\sigma=1$) or antisymmetric ($\sigma=-1$), i.e. 
\begin{equation} \label{Sym-Antisym-Spec}
\phi _{\lambda ^2 \lambda ^1 }^\sigma  \left( {\omega ^2 ,\omega ^1 } \right) = \sigma \phi _{\lambda ^1 \lambda ^2 }^\sigma  \left( {\omega ^1 ,\omega ^2 } \right),
\end{equation}
a condition which also considerably simplifies the mathematical treatment \cite{Symmet}. From Eqs.(\ref{Sym-Antisym-Spec}) and (\ref{ing_wavfun}), the wavefunction corresponding to the spectrum $\phi _{\lambda ^1 \lambda ^2 }^\sigma  \left( {\omega ^1 ,\omega ^2 } \right)$ is
\begin{equation} \label{Sym-Antisym-ing_wavfun}
\Psi _{ss}^{\left( {in} \right) \sigma} \left( {\omega ^1 {\bf{n}}^1 \lambda ^1 ,\omega ^2 {\bf{n}}^2 \lambda ^2 } \right) = \frac{1}{{\sqrt 2 }}\phi _{\lambda ^1 \lambda ^2 }^\sigma  \left( {\omega ^1 ,\omega ^2 } \right)\left[ {\Theta _{{\bf{m}}^1 } \left( {{\bf{n}}^1 } \right)\Theta _{{\bf{m}}^2 } \left( {{\bf{n}}^2 } \right) + \sigma \Theta _{{\bf{m}}^1 } \left( {{\bf{n}}^2 } \right)\Theta _{{\bf{m}}^2 } \left( {{\bf{n}}^1 } \right)} \right]
\end{equation}
so that the exchange ${\bf{m}}^1  \leftrightarrow {\bf{m}}^2$ yields the wavefunction change $\Psi _{ss}^{\left( {in} \right) \sigma}  \to \sigma \Psi _{ss}^{\left( {in} \right) \sigma}$, i.e. it leaves invariant the probability density $| {\Psi _{ss}^{\left( {in} \right)\sigma} } |^2$. Physically, this means that an ingoing wavepacket with symmetric or antisymmetric spectrum is such that the flows of its $s$ polaritons through the two cones around the fixed directions ${\bf m}^1$ and ${\bf m}^2$ shown in Fig.1(a) are identical. It is worth noting that the ingoing $s$ polaritons with antisymmetric spectrum are necessarily entangled since a separable spectrum  $\phi _{\lambda ^1 \lambda ^2 } \left( {\omega ^1 ,\omega ^2 } \right) = \xi _{\lambda ^1 } \left( {\omega ^1 } \right)\zeta _{\lambda ^2 } \left( {\omega ^2 } \right)$ pertaining not-entangled $s$ polaritons can not be antisymmetric; accordingly, the possible detection of interference effects pertinent to the spectral antisymmetry is a distinct signature of the ingoing $s$ polariton entanglement.

Substituting Eq.(\ref{Sym-Antisym-Spec}) into Eqs.(\ref{Psi_ss_se_ee}) and taking the square moduli, we get the outgoing polariton joint detection probability densities
\begin{eqnarray} \label{Prob-Den}
\frac{{dP_{ss}^\sigma  \left( {\omega ^1 {\bf{N}}^1 \Lambda ^1 ,\omega ^2 {\bf{N}}^2 \Lambda ^2 } \right)}}{{d\omega ^1 do_{{\bf{N}}^1 } d\omega ^2 do_{{\bf{N}}^2 } }} &=& \frac{{\Delta o^2 }}{2}\left| {\sum\limits_{\lambda ^1 \lambda ^2 } {\phi _{\lambda ^1 \lambda ^2 }^\sigma \left( {\omega ^1 ,\omega ^2 } \right)\left( {s_{{\bf{m}}^1 \lambda ^1 }^{{\bf{N}}^1 \Lambda ^1 } s_{{\bf{m}}^2 \lambda ^2 }^{{\bf{N}}^2 \Lambda ^2 }  + \sigma s_{{\bf{m}}^1 \lambda ^2 }^{{\bf{N}}^2 \Lambda ^2 } s_{{\bf{m}}^2 \lambda ^1 }^{{\bf{N}}^1 \Lambda ^1 } } \right)} } \right|^2, \nonumber \\ 
 \frac{{dP_{se}^\sigma  \left( {\omega ^1 {\bf{N}}^1 \Lambda ^1 ,\omega ^2 {\bf{R}}^2 \Upsilon ^2 } \right)}}{{d\omega ^1 do_{{\bf{N}}^1 } d\omega ^2 d^3 {\bf{R}}^2 }} &=& \Delta o^2 \left| {\sum\limits_{\lambda ^1 \lambda ^2 } {\phi _{\lambda ^1 \lambda ^2 }^\sigma \left( {\omega ^1 ,\omega ^2 } \right)\left( {s_{{\bf{m}}^1 \lambda ^1 }^{{\bf{N}}^1 \Lambda ^1 } a_{{\bf{m}}^2 \lambda ^2 }^{{\bf{R}}^2 \Upsilon ^2 }  + \sigma a_{{\bf{m}}^1 \lambda ^2 }^{{\bf{R}}^2 \Upsilon ^2 } s_{{\bf{m}}^2 \lambda ^1 }^{{\bf{N}}^1 \Lambda ^1 } } \right)} } \right|^2, \nonumber  \\ 
 \frac{{dP_{ee}^\sigma  \left( {\omega ^1 {\bf{R}}^1 \Upsilon ^1 ,\omega ^2 {\bf{R}}^2 \Upsilon ^2 } \right)}}{{d\omega ^1 d^3 {\bf{R}}^1 d\omega ^2 d^3 {\bf{R}}^2 }} &=& \frac{{\Delta o}}{2}\left| {\sum\limits_{\lambda ^1 \lambda ^2 } {\phi _{\lambda ^1 \lambda ^2 }^\sigma \left( {\omega ^1 ,\omega ^2 } \right)\left( {a_{{\bf{m}}^1 \lambda ^1 }^{{\bf{R}}^1 \Upsilon ^1 } a_{{\bf{m}}^2 \lambda ^2 }^{{\bf{R}}^2 \Upsilon ^2 }  + \sigma a_{{\bf{m}}^1 \lambda ^2 }^{{\bf{R}}^2 \Upsilon ^2 } a_{{\bf{m}}^2 \lambda ^1 }^{{\bf{R}}^1 \Upsilon ^1 } } \right)} } \right|^2  
\end{eqnarray}
whose associated complete scattering description is rather involved due to the number of polariton variables. To simplify the analysis, bearing in mind that outgoing $e$ polaritons are usually left unmeasured in standard setup (i.e. no measurement is performed on the object), we focus on the processes where two, one and zero outgoing $s$ polaritons are globally scattered whose probabilities, in view of the first assumption of Sec.IIIA, are given by 
\begin{eqnarray} \label{P2s_P1s_P0s_01}
 P_{2s} ^\sigma &=& \int {d\omega ^1 } \int {do_{{\bf{N}}^1 } } \sum\limits_{\Lambda ^1 } {\int {d\omega ^2 } \int {do_{{\bf{N}}^2 } } \sum\limits_{\Lambda ^2 } {\frac{{dP_{ss}^\sigma  \left( {\omega ^1 {\bf{N}}^1 \Lambda ^1 ,\omega ^2 {\bf{N}}^2 \Lambda ^2 } \right)}}{{d\omega ^1 do_{{\bf{N}}^1 } d\omega ^2 do_{{\bf{N}}^2 } }}} } , \nonumber \\ 
 P_{1s} ^\sigma &=& \int {d\omega ^1 } \int {do_{{\bf{N}}^1 } } \sum\limits_{\Lambda ^1 } {\int {d\omega ^2 } \int {d^3 {\bf{R}}^2 } \sum\limits_{\Upsilon ^2 } {\frac{{dP_{se}^\sigma  \left( {\omega ^1 {\bf{N}}^1 \Lambda ^1 ,\omega ^2 {\bf{R}}^2 \Upsilon ^2 } \right)}}{{d\omega ^1 do_{{\bf{N}}^1 } d\omega ^2 d^3 {\bf{R}}^2 }}} }, \nonumber  \\ 
 P_{0s} ^\sigma &=& \int {d\omega ^1 } \int {d^3 {\bf{R}}^1 } \sum\limits_{\Upsilon ^1 } {\int {d\omega ^2 } \int {d^3 {\bf{R}}^2 } \sum\limits_{\Upsilon ^2 } {\frac{{dP_{ee}^\sigma  \left( {\omega ^1 {\bf{R}}^1 \Upsilon ^1 ,\omega ^2 {\bf{R}}^2 \Upsilon ^2 } \right)}}{{d\omega ^1 d^3 {\bf{R}}^1 d\omega ^2 d^3 {\bf{R}}^2 }}} }.
\end{eqnarray}
Note that, in the angular integrations of Eqs.(\ref{P2s_P1s_P0s_01}) we have included the outgoing $s$ polariton directions ${\bf N}^1$ and ${\bf N}^2$ belonging to the incident cones of Fig.1(a) (which are excluded by the above first assumption of Sec.IIIA) by commiting a negligible error since $\Delta o \ll 1$. After some algebra, the scattering probabilities of Eqs.(\ref{P2s_P1s_P0s_01}) turn out to be
\begin{eqnarray} \label{P2s_P1s_P0s_02}
P_{2s} ^\sigma &=& \Delta o^2 \left[ \left\langle {S_{{\bf{m}}^1 \lambda '^1 }^{{\bf{m}}^1 \lambda ^1 } S_{{\bf{m}}^2 \lambda '^2 }^{{\bf{m}}^2 \lambda ^2 } } \right\rangle  ^\sigma  + \sigma \left\langle {S_{{\bf{m}}^1 \lambda '^1 }^{{\bf{m}}^2 \lambda ^1 } S_{{\bf{m}}^2 \lambda '^2 }^{{\bf{m}}^1 \lambda ^2 } } \right\rangle  ^\sigma \right], \nonumber  \\ 
P_{1s} ^\sigma &=& \Delta o^2 \left[ \left\langle {S_{{\bf{m}}^1 \lambda '^1 }^{{\bf{m}}^1 \lambda ^1 } A_{{\bf{m}}^2 \lambda '^2 }^{{\bf{m}}^2 \lambda ^2 }  + A_{{\bf{m}}^1 \lambda '^1 }^{{\bf{m}}^1 \lambda ^1 } S_{{\bf{m}}^2 \lambda '^2 }^{{\bf{m}}^2 \lambda ^2 } } \right\rangle  ^\sigma  + \sigma \left\langle {S_{{\bf{m}}^1 \lambda '^1 }^{{\bf{m}}^2 \lambda ^1 } A_{{\bf{m}}^2 \lambda '^2 }^{{\bf{m}}^1 \lambda ^2 }  + A_{{\bf{m}}^1 \lambda '^1 }^{{\bf{m}}^2 \lambda ^1 } S_{{\bf{m}}^2 \lambda '^2 }^{{\bf{m}}^1 \lambda ^2 } } \right\rangle  ^\sigma \right], \nonumber  \\ 
P_{0s} ^\sigma &=& \Delta o^2 \left[ \left\langle {A_{{\bf{m}}^1 \lambda '^1 }^{{\bf{m}}^1 \lambda ^1 } A_{{\bf{m}}^2 \lambda '^2 }^{{\bf{m}}^2 \lambda ^2 } } \right\rangle  ^\sigma  + \sigma \left\langle {A_{{\bf{m}}^1 \lambda '^1 }^{{\bf{m}}^2 \lambda ^1 } A_{{\bf{m}}^2 \lambda '^2 }^{{\bf{m}}^1 \lambda ^2 } } \right\rangle  ^\sigma \right],
\end{eqnarray}
where
\begin{eqnarray} \label{SA}
S_{{\bf{n}}'\lambda '}^{{\bf{n}}\lambda }  \left( \omega  \right) &=& {\bf{e}}_{{\bf{n}}\lambda }  \cdot \left[ \left( {\frac{{k_\omega  }}{{2\pi }}} \right)^2 \int {do_{\bf{N}} } {\cal S}_\omega ^{T*} \left( {\left. {\bf{N}} \right|{\bf{n}}} \right) \cdot {\cal S}_\omega  \left( {\left. {\bf{N}} \right|{\bf{n}}'} \right) \right] \cdot {\bf{e}}_{{\bf{n}}'\lambda '} , \nonumber \\
A_{{\bf{n}}'\lambda '}^{{\bf{n}}\lambda } \left( \omega  \right) &=& {\bf{e}}_{{\bf{n}}\lambda }  \cdot \left[ \int {d^3 {\bf{R}}} {\cal A}_{\omega es}^{T*} \left( {{\bf{R}}\left| {\bf{n}} \right.} \right)  \cdot {\cal A}_{\omega es} \left( {{\bf{R}}\left| {{\bf{n}}'} \right.} \right) \right] \cdot {\bf{e}}_{{\bf{n}}'\lambda '}
\end{eqnarray}
are the dimensionless power scattering and absorption coefficients of the scatterer object satisfying the hermiticity relations $\left( {S_{{\bf{n}}'\lambda '}^{{\bf{n}}\lambda } } \right)^*  = S_{{\bf{n}}\lambda }^{{\bf{n}}'\lambda '}$ and $\left( {A_{{\bf{n}}'\lambda '}^{{\bf{n}}\lambda } } \right)^*  = A_{{\bf{n}}\lambda }^{{\bf{n}}'\lambda '}$ whereas
\begin{align}
\left\langle {Q_{\lambda '^1 \lambda '^2 }^{\lambda ^1 \lambda ^2 } } \right\rangle  ^\sigma  &= \sum\limits_{\lambda ^1 \lambda ^2 \lambda '^1 \lambda '^2 } {\left( I ^\sigma \right)_{\lambda '^1 \lambda '^2 }^{\lambda ^1 \lambda ^2 } } Q_{\lambda '^1 \lambda '^2 }^{\lambda ^1 \lambda ^2 }, &
\left( I ^\sigma \right)_{\lambda '^1 \lambda '^2 }^{\lambda ^1 \lambda ^2 }  &= \int {d\omega ^1 d\omega ^2 } \phi _{\lambda ^1 \lambda ^2 }^{\sigma *} \left( {\omega ^1 ,\omega ^2 } \right)\phi _{\lambda '^1 \lambda '^2 }^\sigma  \left( {\omega ^1 ,\omega ^2 } \right)
\end{align}
is the average of the classical quantity $Q$ performed by means of the spectral overlap integrals $I ^\sigma$ of the ingoing wavepacket spectrum, in turn satisfying the relations (see Eqs.(\ref{spect_norm}) and (\ref{Sym-Antisym-Spec}))
\begin{align} \label{OvIn_prop}
\sum\limits_{\lambda ^1 \lambda ^2 } {\left( I ^\sigma \right)_{\lambda ^1 \lambda ^2 }^{\lambda ^1 \lambda ^2 } }  &= 1, &
\left[ {\left( I ^\sigma \right)_{\lambda '^1 \lambda '^2 }^{\lambda ^1 \lambda ^2 } } \right]^* &=  \left( I ^\sigma \right)_{\lambda ^1 \lambda ^2 }^{\lambda '^1 \lambda '^2 } , & 
\left( I ^\sigma \right)_{\lambda '^2 \lambda '^1 }^{\lambda ^2 \lambda ^1 }  &= \left( I ^\sigma \right)_{\lambda '^1 \lambda '^2 }^{\lambda ^1 \lambda ^2 } . 
\end{align}
Note that three probabilities in Eqs.(\ref{P2s_P1s_P0s_02}) have two contributions, respectively  characterized by their ${\bf m}^1{\bf m}^2$ label placements $\left( {{}_{{\bf{m}}^1 }^{{\bf{m}}^1 } {}_{{\bf{m}}^2 }^{{\bf{m}}^2 } } \right)$ and $\left( {{}_{{\bf{m}}^1 }^{{\bf{m}}^2 } {}_{{\bf{m}}^2 }^{{\bf{m}}^1 } } \right)$, the second of which have the factor $\sigma$. From all the above discussion, it follows that the $\left( {{}_{{\bf{m}}^1 }^{{\bf{m}}^2 } {}_{{\bf{m}}^2 }^{{\bf{m}}^1 } } \right)$ terms are responsible for the quantum interference effects showing up in the $s$ polariton scattering probabilities and that such interference is affected by the symmetry $\sigma$ of the ingoing wavepacket spectrum.

\subsection{Physical meaning of the power scattering and absorption coefficients}
Before undertaking the analysis of the wavepacket quantum scattering by a sphere, we here discuss the physical meaning of the power scattering and absorption coefficients in Eqs.(\ref{SA}). To this end we consider the classical scattering by an arbitrary lossy object of the incindent field ${\bf{E}}_\omega ^{\left( {in} \right)} \left( {\bf{s}} \right) = E_0 \sum\nolimits_{i = 1}^N {e^{ik_\omega  {\bf{m}}^i  \cdot {\bf{s}}} {\bf{e}}^i }$ which is the superposition of $N$ plane waves with equal amplitude $E_0$, directions ${\bf{m}}^i$ and polarizations ${\bf{e}}^i$ (with ${\bf{m}}^i  \cdot {\bf{m}}^i  = 1$, ${\bf{e}}^i  \cdot {\bf{e}}^{i*}  = 1$ and
${{\bf{e}}^i  \cdot {\bf{m}}^i  = 0}$). A simple extension of the single plane wave scattering analysis (e.g. see  Ref. \cite{Krist1}), enables to prove that the $N$ planes scattering is characterized by the scattering, extinction and absorption cross-sections given by
\begin{eqnarray}
 C_{sca}  &=& \sum\limits_{ij} {{\bf{e}}^{i*}  \cdot \left[ {\int {do_{\bf{N}} } {\cal S}_\omega ^{T*} \left( {{\bf{N}}\left| {{\bf{m}}^i } \right.} \right) \cdot {\cal S}_\omega  \left( {{\bf{N}}\left| {{\bf{m}}^j } \right.} \right)} \right] \cdot {\bf{e}}^j }, \nonumber \\
 C_{ext}  &=& \sum\limits_{ij} {{\bf{e}}^{i*}  \cdot \frac{{2\pi }}{{ik_\omega  }}\left[ {{\cal S}_\omega  \left( {{\bf{m}}^i \left| {{\bf{m}}^j } \right.} \right) - {\cal S}_\omega ^{T*} \left( {{\bf{m}}^j \left| {{\bf{m}}^i } \right.} \right)} \right] \cdot {\bf{e}}^j }, \nonumber \\
 C_{abs}  &=& C_{ext}  - C_{sca},
\end{eqnarray}
where each cross-section is the ratio between the corresponding power and the intensity $\left| {E_0 } \right|^2 /\left( {2\mu _0 c} \right)$ of each plane wave. It is essential now to observe that, using the transmission dyadic ${{\cal T}_{\omega ss} \left( {\left. {\bf{N}} \right|{\bf{n}}} \right)}$ in the first of Eqs.(\ref{QOS_ClasDyad}), the first of Eqs.(\ref{QOS_EnBal1}) turns into
\begin{equation} \label{Fund_Ene_Cons}
\int {d^3 {\bf{R}}} {\cal A}_{\omega es}^{T*} \left( {{\bf{R}}\left| {\bf{n}} \right.} \right) \cdot {\cal A}_{\omega es} \left( {{\bf{R}}\left| {{\bf{n}}'} \right.} \right) = \left( {\frac{{k_\omega  }}{{2\pi }}} \right)^2 \left\{ {\frac{{2\pi }}{{ik_\omega  }}\left[ {{\cal S}_\omega  \left( {\left. {\bf{n}} \right|{\bf{n}}'} \right) - {\cal S}_\omega ^{T*} \left( {\left. {{\bf{n}}'} \right|{\bf{n}}} \right)} \right] - \int {do_{\bf{N}} } {\cal S}_\omega ^{T*} \left( {\left. {\bf{N}} \right|{\bf{n}}} \right) \cdot {\cal S}_\omega  \left( {\left. {\bf{N}} \right|{\bf{n}}'} \right)} \right\}
\end{equation}
whose straightforward consequence is that the absorption cross-section can be rewritten as
\begin{equation}
C_{abs}  = \left( {\frac{{2\pi }}{{k_\omega  }}} \right)^2 \sum\limits_{ij} {{\bf{e}}^{i*}  \cdot } \left[ {\int {d^3 {\bf{R}}} {\cal A}_{\omega es}^{T*} \left( {{\bf{R}}\left| {{\bf{m}}^i } \right.} \right) \cdot {\cal A}_{\omega es} \left( {{\bf{R}}\left| {{\bf{m}}^j } \right.} \right)} \right] \cdot {\bf{e}}^j.
\end{equation}
Eventually, by expanding the polarizations ${{\bf{e}}^i }$ of the plane waves on the hereafter considered basis $\left\{ {{\bf{e}}_{{\bf{m}}^i 1} ,{\bf{e}}_{{\bf{m}}^i 2} } \right\}$, i.e. setting ${\bf{e}}^i  = \sum\nolimits_\lambda  {c_\lambda ^i } {\bf{e}}_{{\bf{m}}^i \lambda }$, and using Eqs.(\ref{SA}), the scattering and absorption cross-section can be cast as
\begin{align} \label{Gen_Csca_Cabs}
 C_{sca}  &= \left( {\frac{{2\pi }}{{k_\omega  }}} \right)^2 \sum\limits_{ij} {\sum\limits_{\lambda \lambda '} {c_\lambda ^{i*} c_{\lambda '}^j S_{{\bf{m}}^j \lambda '}^{{\bf{m}}^i \lambda } } }, &
 C_{abs}  &= \left( {\frac{{2\pi }}{{k_\omega  }}} \right)^2 \sum\limits_{ij} {\sum\limits_{\lambda \lambda '} {c_\lambda ^{i*} c_{\lambda '}^j A_{{\bf{m}}^j \lambda '}^{{\bf{m}}^i \lambda } } },
\end{align}
expressions elucidating the physical meaning of the power scattering and absorption coefficients. Specifically, apart from the numerical factor $\left( {2\pi /k_\omega  } \right)^2$, the diagonal terms $S_{{\bf{n}}\lambda }^{{\bf{n}}\lambda }$ and $A_{{\bf{n}}\lambda }^{{\bf{n}}\lambda }$ represent the usual scattering and absorption cross sections of the plane wave $e^{ik_\omega  {\bf{n}} \cdot {\bf{s}}} {\bf{e}}_{{\bf{n}}\lambda } $ whereas the off-diagonal terms $S_{{\bf{n}}'\lambda '}^{{\bf{n}}\lambda }$ and $A_{{\bf{n}}'\lambda '}^{{\bf{n}}\lambda }$ represent the contributions of the interference bewteen the plane waves $e^{ik_\omega  {\bf{n}} \cdot {\bf{s}}} {\bf{e}}_{{\bf{n}}\lambda }$ and $e^{ik_\omega  {\bf{n}}' \cdot {\bf{s}}} {\bf{e}}_{{\bf{n}}'\lambda '}$ to the overall scattering and absorption cross sections.

\subsection{Homogeneous lossy sphere}
We now specialize the above general description of quantum scattering to the relevant and analytically treatable situation where the scatterer object is a homogeneous lossy sphere of radius $a$ and uniform dielectric permittivity $\varepsilon _\omega ^V$. Since we will be mainly interested in analyzing the coincidence detection and survival probabilities of the outgoing $s$ polaritons (see the First of Eqs.(\ref{Psi_ss_se_ee}) and Eqs.(\ref{P2s_P1s_P0s_02})), we here discuss the scattering coefficient $s_{{\bf{n}}\lambda }^{{\bf{N}}\Lambda }$ in  Eqs.(\ref{sa}) and the power scattering and absorption coefficients $S_{{\bf{n}}'\lambda '}^{{\bf{n}}\lambda }  $ and $A_{{\bf{n}}'\lambda '}^{{\bf{n}}\lambda }$ in Eqs.(\ref{SA}) of a homogeneous sphere, which (as detailed in Appendix A) are given by   
\begin{eqnarray} \label{Sph_sSA}
 s_{{\bf{n}}\lambda }^{{\bf{N}}\Lambda } \left( \omega  \right) &=& {\rm J}_{{\bf{N}} \cdot {\bf{n}}} \left\{ {\begin{pmatrix}
   { - \frac{1}{2}a_{n\omega } } & { - \frac{1}{2}b_{n\omega } }  \\
\end{pmatrix}} \right\}\begin{pmatrix}
   {{\bf{e}}_{{\bf{N}}\Lambda }  \cdot {\bf{e}}_{{\bf{n}}\lambda } }  \\
   {{\bf{e}}_{{\bf{N}}\Lambda }  \cdot {\bf{nN}} \cdot {\bf{e}}_{{\bf{n}}\lambda } }  \\
\end{pmatrix}, \nonumber \\ 
 S_{{\bf{n}}'\lambda '}^{{\bf{n}}\lambda } \left( \omega  \right) &=& {\rm J}_{{\bf{n}} \cdot {\bf{n}}'} \left\{ {\begin{pmatrix}
   {\left| {a_{n\omega } } \right|^2 } & {\left| {b_{n\omega } } \right|^2 }  \\
\end{pmatrix}} \right\}\begin{pmatrix}
   {{\bf{e}}_{{\bf{n}}\lambda }  \cdot {\bf{e}}_{{\bf{n}}'\lambda '} }  \\
   {{\bf{e}}_{{\bf{n}}\lambda }  \cdot {\bf{n}}'{\bf{n}} \cdot {\bf{e}}_{{\bf{n}}'\lambda '} }  \\
\end{pmatrix}, \nonumber \\ 
 A_{{\bf{n}}'\lambda '}^{{\bf{n}}\lambda } \left( \omega  \right) &=& {\rm J}_{{\bf{n}} \cdot {\bf{n}}'} \left\{ {\begin{pmatrix}
   {{\mathop{\rm Re}\nolimits} \left( {a_{n\omega } } \right) - \left| {a_{n\omega } } \right|^2 } & {{\mathop{\rm Re}\nolimits} \left( {b_{n\omega } } \right) - \left| {b_{n\omega } } \right|^2 }  \\
\end{pmatrix}} \right\}\begin{pmatrix}
   {{\bf{e}}_{{\bf{n}}\lambda }  \cdot {\bf{e}}_{{\bf{n}}'\lambda '} }  \\
   {{\bf{e}}_{{\bf{n}}\lambda }  \cdot {\bf{n}}'{\bf{n}} \cdot {\bf{e}}_{{\bf{n}}'\lambda '} }  \\
\end{pmatrix},
\end{eqnarray}
where, for any pair of complex sequences $p_n$ and $q_m$, we have introduced the row vector
\begin{equation} \label{O}
{\rm J}_\xi  \left\{ {\begin{pmatrix}
   {p_n } & {q_m }  \\
\end{pmatrix}} \right\} = \frac{1}{\pi }\sum\limits_{n = 1}^\infty  {\frac{{2n + 1}}{{n\left( {n + 1} \right)}}} \begin{pmatrix}
   {p_n } & {q_m }  \\
\end{pmatrix}\begin{pmatrix}
   {P'_n \left( \xi  \right)} & {P''_n \left( \xi  \right)}  \\
   { - \left( {1 - \xi ^2 } \right)P''_n \left( \xi  \right) + \xi P'_n \left( \xi  \right)} & { - \xi P''_n \left( \xi  \right) - P'_n \left( \xi  \right)}  \\
\end{pmatrix},
\end{equation}
where $P_n \left( \xi  \right)$ is the Legendre polynomial and $a_{n\omega } , b_{n\omega }$ are the standard Mie coefficients pertaining the scattering of a single plane wave by a homogeneous sphere \cite{Bohre}, i.e.
\begin{eqnarray} \label{Sph_anbn}
 a_{n\omega }  &=& \frac{\displaystyle {j_n \left( {k_\omega ^V a} \right)\frac{\partial }{{\partial a}}\left[ {aj_n \left( {k_\omega  a} \right)} \right] - \frac{1}{{\varepsilon _\omega ^V }}j_n \left( {k_\omega  a} \right)\frac{\partial }{{\partial a}}\left[ {aj_n \left( {k_\omega ^V a} \right)} \right]}}{\displaystyle {j_n \left( {k_\omega ^V a} \right)\frac{\partial }{{\partial a}}\left[ {ah_n^{\left( 1 \right)} \left( {k_\omega  a} \right)} \right] - \frac{1}{{\varepsilon _\omega ^V }}h_n^{\left( 1 \right)} \left( {k_\omega  a} \right)\frac{\partial }{{\partial a}}\left[ {aj_n \left( {k_\omega ^V a} \right)} \right]}}, \nonumber \\ 
 b_{n\omega }  &=& \frac{\displaystyle{j_n \left( {k_\omega ^V a} \right)\frac{\partial }{{\partial a}}\left[ {aj_n \left( {k_\omega  a} \right)} \right] - j_n \left( {k_\omega  a} \right)\frac{\partial }{{\partial a}}\left[ {aj_n \left( {k_\omega ^V a} \right)} \right]}}{\displaystyle{j_n \left( {k_\omega ^V a} \right)\frac{\partial }{{\partial a}}\left[ {ah_n^{\left( 1 \right)} \left( {k_\omega  a} \right)} \right] - h_n^{\left( 1 \right)} \left( {k_\omega  a} \right)\frac{\partial }{{\partial a}}\left[ {aj_n \left( {k_\omega ^V a} \right)} \right]}},
\end{eqnarray}
where $k_\omega ^V  = k_\omega  \sqrt {\varepsilon _\omega ^V } $ (with ${{\mathop{\rm Im}\nolimits} ( {k_\omega ^V } ) \ge 0}$) is the wavenumber inside the sphere and $j_n \left( \rho  \right)$ and $h_n^{\left( 1 \right)} \left( \rho  \right)$ are the spherical Bessel and Hankel functions of the first kind. 

Note that the direction ${\bf{m}} = \left\{ {{\bf{N}},{\bf{n}},{\bf{n}}'} \right\}$ and polarization ${\bf{e}}_{{\bf{m}}\mu }  = \left\{ {{\bf{e}}_{{\bf{N}}\Lambda } ,{\bf{e}}_{{\bf{n}}\lambda } ,{\bf{e}}_{{\bf{n}}'\lambda '} } \right\}$ unit vectors appear  in Eqs.(\ref{Sph_sSA}) only inside scalar products, with the result that  $s_{{\bf{n}}\lambda }^{{\bf{N}}\Lambda }$, $S_{{\bf{n}}'\lambda '}^{{\bf{n}}\lambda }$ and $A_{{\bf{n}}'\lambda '}^{{\bf{n}}\lambda }$ are left invariant by any proper or improper rotation of the whole scattering setup ${\bf{m}} \to {\cal R} \cdot {\bf{m}}$ and ${\bf{e}}_{{\bf{m}}\mu }  \to {\cal R} \cdot {\bf{e}}_{{\bf{m}}\mu }$  (where ${\cal R}$ is an orthogonal dyadic i.e. ${\cal R} \cdot {\cal R}^T  = {\cal R}^T  \cdot {\cal R} = {\cal I}$), an evident consequence of the sphere symmetry. From Eqs.(\ref{Sph_sSA}), the diagonal power scattering and absorption coefficients are
\begin{align}
S_{{\bf{n}}\lambda }^{{\bf{n}}\lambda } \left( \omega  \right) &= {\rm J}_1 \left\{ {\begin{pmatrix}
   {\left| {a_{n\omega } } \right|^2 } & {\left| {b_{n\omega } } \right|^2 }  \\
\end{pmatrix}} \right\}\begin{pmatrix}
   1  \\
   0  \\
\end{pmatrix}, &
A_{{\bf{n}}\lambda }^{{\bf{n}}\lambda } \left( \omega  \right) &= {\rm J}_1 \left\{ {\begin{pmatrix}
   {{\mathop{\rm Re}\nolimits} \left( {a_{n\omega } } \right) - \left| {a_{n\omega } } \right|^2 } & {{\mathop{\rm Re}\nolimits} \left( {b_{n\omega } } \right) - \left| {b_{n\omega } } \right|^2 }  \\
\end{pmatrix}} \right\}\begin{pmatrix}
   1  \\
   0  \\
\end{pmatrix},
\end{align}
so that, since $P'_n \left( 1 \right) = n\left( {n + 1} \right)/2$, Eq.(\ref{O}) directly yields
\begin{eqnarray} \label{Sphere_CscaSabs}
 \left( {\frac{{2\pi }}{{k_\omega  }}} \right)^2 S_{{\bf{n}}\lambda }^{{\bf{n}}\lambda } \left( \omega  \right) &&= \frac{{2\pi }}{{k_\omega ^2 }}\sum\limits_{n = 1}^\infty  {\left( {2n + 1} \right)} \left( {\left| {a_{n\omega } } \right|^2  + \left| {b_{n\omega } } \right|^2 } \right) \equiv C_{sca}^S, \nonumber  \\ 
 \left( {\frac{{2\pi }}{{k_\omega  }}} \right)^2 A_{{\bf{n}}\lambda }^{{\bf{n}}\lambda } \left( \omega  \right) &=& \frac{{2\pi }}{{k_\omega ^2 }}\sum\limits_{n = 1}^\infty  {\left( {2n + 1} \right)\left\{ {\left[ {{\mathop{\rm Re}\nolimits} \left( {a_{n\omega } } \right) - \left| {a_{n\omega } } \right|^2 } \right] + \left[ {{\mathop{\rm Re}\nolimits} \left( {b_{n\omega } } \right) - \left| {b_{n\omega } } \right|^2 } \right]} \right\} \equiv } C_{abs}^S,
\end{eqnarray}
where $C_{sca}^S$ and $C_{abs}^S$ are the classical scattering and absorption cross-sections pertaining the scattering of any single plane wave $e^{ik_\omega  {\bf{n}} \cdot {\bf{s}}} {\bf{e}}_{{\bf{n}}\lambda }$ by the sphere \cite{Bohre}, in agreement with the discussion in Sec.IIIC. As a last remark on the coefficients in Eqs.(\ref{Sph_sSA}), it is worth discussing the two specific situations where there is no sphere $\varepsilon _\omega ^V  = 1$ or the sphere is ideally transparent ${\mathop{\rm Im}\nolimits} ( {\varepsilon _\omega ^V } ) = 0$, they providing consistency checks of our treatment. In the first situation the Mie coefficients $a_{n\omega }$ and $b_{n\omega }$ vanish (see Eq.(\ref{Sph_anbn})) and consequently, from Eqs.(\ref{Sph_sSA}), the coefficients $s_{{\bf{n}}\lambda }^{{\bf{N}}\Lambda }$, $S_{{\bf{n}}'\lambda '}^{{\bf{n}}\lambda }$ and $A_{{\bf{n}}'\lambda '}^{{\bf{n}}\lambda }$ vanish as well, i.e. there is correctly no radiation scattering. In the second situation, as shown in Appendix A, the Mie coefficients are such that ${\mathop{\rm Re}\nolimits} \left( {a_{n\omega } } \right) = \left| {a_{n\omega } } \right|^2$ and ${\mathop{\rm Re}\nolimits} \left( {b_{n\omega } } \right) = \left| {b_{n\omega } } \right|^2$ so that, from the third of Eqs.(\ref{Sph_sSA}), the power absorption coefficient $A_{{\bf{n}}'\lambda '}^{{\bf{n}}\lambda }$ vanishes, i.e the sphere correctly does not absorb energy from the radiation.

\begin{figure}
\centering
\includegraphics[width = 1\linewidth]{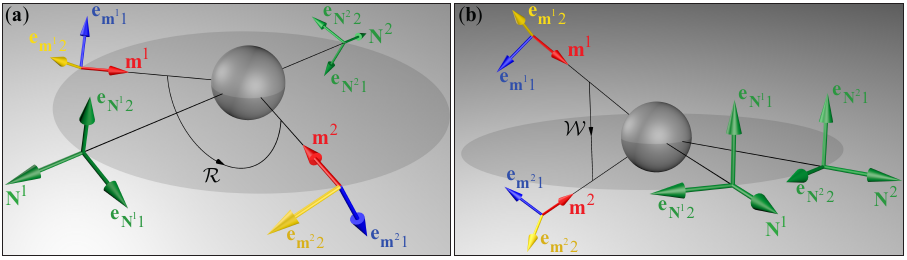}
\caption{Sample classes of scattering geometries leading to perfect constructive or destructive interference in the polariton coincidence detection. The ingoing polariton triads $\left\{ {{\bf{m}}^1 ,{\bf{e}}_{{\bf{m}}^1 1} ,{\bf{e}}_{{\bf{m}}^1 2} } \right\}$ and $\left\{ {{\bf{m}}^2 ,{\bf{e}}_{{\bf{m}}^2 1} ,{\bf{e}}_{{\bf{m}}^2 2} } \right\}$, representing the input ports through which light is launched, are plotted with colors outlining the action of the rotation ${\cal R}$ and the reflection ${\cal W}$  whereas the outgoing polariton triads $\left\{ {{\bf{N}}^1 ,{\bf{e}}_{{\bf{N}}^1 1} ,{\bf{e}}_{{\bf{N}}^1 2} } \right\}$ and $\left\{ {{\bf{N}}^2 ,{\bf{e}}_{{\bf{N}}^2 1} ,{\bf{e}}_{{\bf{N}}^2 2} } \right\}$, representing the detectors spatial placements and polarization directions, are plotted in green.}
\label{Fig1}
\end{figure}

\section{Scattering geometries for perfect interference in the polariton coincidence detection}
In addition to the overall rotational invariance mentioned in Section.IIID, we here show that the sphere symmetry entails the existence of specific scattering geometries such that, regardless of the wavepacket spectral profile, the probability density of polariton coincidence detection  displays perfect constructive or destructive  interference.

With reference to the $ss \to ss$ process discussed in Sec.IIIA, a scattering geometry is assigned by the four orthonormal triads $\left\{ {{\bf{m}}^1 ,{\bf{e}}_{{\bf{m}}^1 1}, {\bf{e}}_{{\bf{m}}^1 2} } \right\}$, $\left\{ {{\bf{m}}^2,{\bf{e}}_{{\bf{m}}^2 1} ,{\bf{e}}_{{\bf{m}}^2 2} } \right\}$, $\left\{ {{\bf{N}}^1 ,{\bf{e}}_{{\bf{N}}^1 1} ,{\bf{e}}_{{\bf{N}}^1 2} } \right\}$, $\left\{ {{\bf{N}}^2 ,{\bf{e}}_{{\bf{N}}^2 1} ,{\bf{e}}_{{\bf{N}}^2 2} } \right\}$, i.e. by the directions and polarizations of the ingoing and outgoing $s$ polaritons. In other words, a scattering geometry represents the geometric arrangement is space of the 'input ports' through which light is launched and of the 'output ports' hosting the detectors, regardless of the spectrum $\phi ^\sigma$ accounting for the quantum correlations between the ingoing $s$-polaritons. Now the first of Eqs.(\ref{Prob-Den}) shows that the probability density of polariton coincidence detection depends on the scattering geometry only through the terms ${s_{{\bf{m}}^1 \lambda ^1 }^{{\bf{N}}^1 \Lambda ^1 } s_{{\bf{m}}^2 \lambda ^2 }^{{\bf{N}}^2 \Lambda ^2 } }$ and ${s_{{\bf{m}}^1 \lambda ^2 }^{{\bf{N}}^2 \Lambda ^2 } s_{{\bf{m}}^2 \lambda ^1 }^{{\bf{N}}^1 \Lambda ^1 } }$ representing the electromagnetic contributions to the two quantum paths involved in the $ss \to ss$ process, as discussed in Sec.IIIA. Therefore, a geometric strategy to achieve perfect interference involves exploiting the spherical symmetry to make the two quantum paths fully equivalent or, in other words, to identify the scattering geometries such that the relation
\begin{equation} \label{Per_int}
\sum\limits_{\lambda ^1 \lambda ^2 } {\phi _{\lambda ^1 \lambda ^2 }^\sigma  \left( {\omega ^1 ,\omega ^2 } \right)s_{{\bf{m}}^1 \lambda ^2 }^{{\bf{N}}^2 \Lambda ^2 } s_{{\bf{m}}^2 \lambda ^1 }^{{\bf{N}}^1 \Lambda ^1 } }  = \pm \sum\limits_{\lambda ^1 \lambda ^2 } {\phi _{\lambda ^1 \lambda ^2 }^\sigma  \left( {\omega ^1 ,\omega ^2 } \right)s_{{\bf{m}}^1 \lambda ^1 }^{{\bf{N}}^1 \Lambda ^1 } s_{{\bf{m}}^2 \lambda ^2 }^{{\bf{N}}^2 \Lambda ^2 } } 
\end{equation}
holds. Accordingly, Eq.(\ref{Per_int}) causes  the probability density of polariton coincidence detection in the first of Eqs.(\ref{Prob-Den}) to have the overall factor $\left( {1 \pm \sigma } \right)^2$ (see below) so that perfect interferece is achieved independetly of the spectrum frequency profile. Due to the high symmetry of the sphere, there are many classes of scattering geometries ensuring the validity of Eq.(\ref{Per_int}) but, for conciseness, we here discuss only two main examples. 

The first class we consider contains the scattering geometries, sketched in Fig.2(a), whose outgoing polariton triads have ${\bf{N}}^2  =  - {\bf{N}}^1$ and arbitrary ${\bf{e}}_{{\bf{N}}^i \Lambda }$ ($i=1,2$) and whose ingoing polariton triads are mutually related by a $\pi$ rotation around ${\bf{N}}^i$, i.e. ${\bf{m}}^2  = {\cal R}  \cdot {\bf{m}}^1$ and ${\bf{e}}_{{\bf{m}}^2 \lambda }  = {\cal R} \cdot {\bf{e}}_{{\bf{m}}^1 \lambda }$ where ${\cal R} =  - {\cal I} + 2{\bf{N}}^i {\bf{N}}^i$ is the rotation dyadic. Such scattering geometries correspond to actual setups where the two detectors, with arbitrary polarization directions, are symmetrically aligned around the sphere, whereas the two input ports through which the ingoing $s$ polaritons are launched are mutually related by a half-circle rotation around the detectors direction. The virtue of such scattering geometries is that the two outgoing polariton triads are very easily transformed by the above rotation as ${\cal R} \cdot {\bf{N}}^i  = {\bf{N}}^i$ and ${\cal R} \cdot {\bf{e}}_{{\bf{N}}^i \Lambda }  =  - {\bf{e}}_{{\bf{N}}^i \Lambda }$ so that, from the first of Eqs.(\ref{Sph_sSA}), we readily get the relation $s_{{\bf{m}}^2 \lambda }^{{\bf{N}}^i \Lambda }  =  - s_{{\bf{m}}^1 \lambda }^{{\bf{N}}^i \Lambda }$. Accordingly, Eq.(\ref{Per_int}) holds with the $+$ sign for any wavepacket spectrum ${\phi _{\lambda ^1 \lambda ^2 }^\sigma  \left( {\omega ^1 ,\omega ^2 } \right)}$ and the probability density of polariton coincidence detection in the first of Eqs.(\ref{Prob-Den}) becomes
\begin{equation} \label{Per_int_prob_den}
\frac{{dP_{ss}^\sigma  \left( {\omega ^1 {\bf{N}}^1 \Lambda ^1 ,\omega ^2 {\bf{N}}^2 \Lambda ^2 } \right)}}{{d\omega ^1 do_{{\bf{N}}^1 } d\omega ^2 do_{{\bf{N}}^2 } }} = \left( {1 + \sigma } \right)^2 \frac{{\Delta o^2 }}{2}\left| {\sum\limits_{\lambda ^1 \lambda ^2 } {\phi _{\lambda ^1 \lambda ^2 }^\sigma  \left( {\omega ^1 ,\omega ^2 } \right)s_{{\bf{m}}^1 \lambda ^1 }^{{\bf{N}}^1 \Lambda ^1 } s_{{\bf{m}}^2 \lambda ^2 }^{{\bf{N}}^2 \Lambda ^2 } } } \right|^2.
\end{equation}
We conclude that, in a scattering geometry of this class, for any kind of ingoing $s$ polaritons correlation, the coincidence detection of two outgoing polaritons of any frequency and any polarization shows perfect constructive or destructive interference according to whether the spectrum is symmetric $\sigma = 1$ or antisymmetric $\sigma = -1$, respectively.

The second class we consider contains the scattering geometries, sketched in Fig.2(b), whose outgoing polariton triads have arbitrary ${\bf{N}}^i$ ($i=1,2$), ${\bf{e}}_{{\bf{N}}^1 1}  = {\bf{e}}_{{\bf{N}}^2 1}$ orthogonal to the ${\bf{N}}^1 {\bf{N}}^2$ plane and ${\bf{e}}_{{\bf{N}}^i 2}  = {\bf{N}}^i  \times {\bf{e}}_{{\bf{N}}^i 1}$ and whose ingoing polariton triads are mutually related by a reflection through the ${\bf{N}}^1 {\bf{N}}^2$ plane, i.e. ${\bf{m}}^2  = {\cal W}  \cdot {\bf{m}}^1$ and ${\bf{e}}_{{\bf{m}}^2 \lambda }  = {\cal W} \cdot {\bf{e}}_{{\bf{m}}^1 \lambda }$ where ${\cal W} = {\cal I} - 2{\bf{e}}_{{\bf{N}}^i 1} {\bf{e}}_{{\bf{N}}^i 1}$ is the reflection dyadic.  In an actual setup corresponding to a scattering geometry of this class, the two detectors can be arbitrarily placed in space with their polarization direction either belonging to or orthogonal to the detectors plane, whereas the two input ports through which the ingoing $s$ polaritons are lauched are specular imgages of each other with respect to the detectors plane. In a scattering geometry of this class, the two outgoing polariton triads are transformed by the above reflectection as ${\cal W} \cdot {\bf{N}}^i  = {\bf{N}}^i$ and ${\cal W} \cdot {\bf{e}}_{{\bf{N}}^i \Lambda }  = \left( { - \delta _{\Lambda 1}  + \delta _{\Lambda 2} } \right){\bf{e}}_{{\bf{N}}^i \Lambda }$ so that the first of Eqs.(\ref{Sph_sSA}) readly implies that $s_{{\bf{m}}^2 \lambda }^{{\bf{N}}^i \Lambda }  = \left( { - \delta _{\Lambda 1}  + \delta _{\Lambda 2} } \right)s_{{\bf{m}}^1 \lambda }^{{\bf{N}}^i \Lambda }$. Therefore, Eq.(\ref{Per_int}) holds with sign $\left( { - \delta _{\Lambda ^1 1}  + \delta _{\Lambda ^1 2} } \right)\left( { - \delta _{\Lambda ^2 1}  + \delta _{\Lambda ^2 2} } \right) 
 = 2\delta _{\Lambda ^1 \Lambda ^2 }  - 1$ for any wavepacket spectrum ${\phi _{\lambda ^1 \lambda ^2 }^\sigma  \left( {\omega ^1 ,\omega ^2 } \right)}$ and the probability density of polariton coincidence detection in the first of Eqs.(\ref{Prob-Den}) becomes
\begin{equation}
\frac{{dP_{ss}^\sigma  \left( {\omega ^1 {\bf{N}}^1 \Lambda ^1 ,\omega ^2 {\bf{N}}^2 \Lambda ^2 } \right)}}{{d\omega ^1 do_{{\bf{N}}^1 } d\omega ^2 do_{{\bf{N}}^2 } }} = \left[ {1 + \sigma \left( {2\delta _{\Lambda ^1 \Lambda ^2 }  - 1} \right)} \right]^2 \frac{{\Delta o^2 }}{2}\left| {\sum\limits_{\lambda ^1 \lambda ^2 } {\phi _{\lambda ^1 \lambda ^2 }^\sigma  \left( {\omega ^1 ,\omega ^2 } \right)s_{{\bf{m}}^1 \lambda ^1 }^{{\bf{N}}^1 \Lambda ^1 } s_{{\bf{m}}^2 \lambda ^2 }^{{\bf{N}}^2 \Lambda ^2 } } } \right|^2.
\end{equation}
In plain English, in a scattering geometry of this class, for any kind of ingoing $s$ polaritons correlation, the coincidence detection of two outgoing polaritons of any frequency and with same polarization type $\Lambda^1 = \Lambda^2$ shows perfect constructive or destructive interference according to whether the spectrum is symmetric $\sigma = 1$ or antisymmetric $\sigma = -1$, respectively. Conversely, the coincidence detection of two outgoing polaritons of any frequency and with opposite polarization types $\Lambda^1 \neq \Lambda^2$ shows perfect constructive or destructive interference according to whether the spectrum is antisymmetric $\sigma = -1$ or symmetric $\sigma = 1$, respectively.

It is worth comparing such results with two-photon interference in a $50/50$ lossles beam splitter \cite{Wangg1} which is well known to yield perfect destructive interference or Hong-Ou-Mandel effect (photon coalescence) and perfect constructive interference (photon anticoalescence) according to whether the ingoing two-photon state is symmetric or antisymmetric, respectively. A first striking difference is that the key equality of the transmission and reflection rates is the only symmetry requirement enabling the beam splitter to yield perfect interference whereas the sphere, due to its uncountable number of input/output ports and higher symmetry, dramatically admits a plenty of suitable scattering geometries, as discussed above. Second the beam splitter transparency requirement uniquely links perfect constructive and destructive interference to symmetric and antisymmetic ingoing two-photon states, respectively, whereas in the case of the sphere the relation between perfect interference type and ingoing spectrum symmetry generally depends on the selected scattering geometry and the chosen detection polarizations.

\begin{figure}
\centering
\includegraphics[width = 0.5\linewidth]{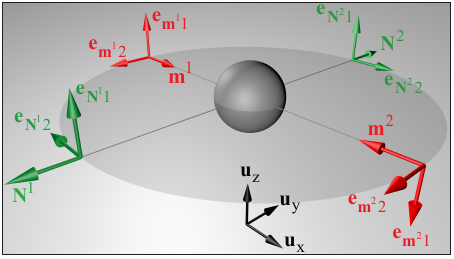}
\caption{Selected scattering geometry for discussing the spectral features of quantum interference effects.}
\label{Fig3}
\end{figure}

\section{Spectral features of quantum interference effects}
The major virtue of the quantum optical scattering approach (and of the MLNF) reviewed in Sec.II is that it enables to investigate the impact of macroscopic matter dispersion/absorption on quantum phenomena. Accordingly, we here model the sphere permittivity by means of the Drude-Lorentz model $\varepsilon _\omega ^V  = 1 + \omega _p^2 /\left( {\omega _0^2  - \omega ^2  - i\omega \gamma } \right)$ and we analyze the impact of the sphere dispersion on the quantum interference effects pertaining the polariton coincidence detection and global scattering processes. In order to both reduce the analysis complexity and to highlight the interference effects, we select the specific scattering geometry, sketched in Fig.3, whose ingoing and outgoing polariton triads are ${\bf{m}}^i = \left( {\delta _{i1}  - \delta _{i2} } \right){\bf{u}}_x$, ${\bf{e}}_{{\bf{m}}^i \lambda }  = \left( {\delta _{i1}  - \delta _{i2} } \right)\delta _{\lambda 1} {\bf{u}}_z  - \delta _{\lambda 2} {\bf{u}}_y$ and ${\bf{N}}^i  = \left( { - \delta _{i1}  + \delta _{i2} } \right){\bf{u}}_y$, ${\bf{e}}_{{\bf{N}}^i \Lambda }  = \delta _{\Lambda 1} {\bf{u}}_z  + \left( { - \delta _{i1}  + \delta _{i2} } \right)\delta _{\Lambda 2} {\bf{u}}_x$. 

By using  the Legendre polynomials properties $P'_n \left( 0 \right) = nP_{n - 1} \left( 0 \right)$, $P''_n \left( 0 \right) =  - n\left( {n + 1} \right)P_n \left( 0 \right)$ and $P_{2m + 1} \left( 0 \right) = 0$, the scattering coefficient in the first of Eqs.(\ref{Sph_sSA}) for such scattering geometry turns out to be
\begin{equation}
s_{{\bf{m}}^j \lambda }^{{\bf{N}}^i \Lambda } \left( \omega  \right) = \left( { - \delta _{j1}  + \delta _{j2} } \right)\left( {\delta _{\Lambda 1} \delta _{\lambda 1} T_\omega ^{oe}  + \delta _{\Lambda 2} \delta _{\lambda 2} T_\omega ^{eo} } \right)
\end{equation}
where
\begin{eqnarray} \label{Disp_ToeTeo}
 T_\omega ^{oe}  &=&  \frac{1}{{2\pi }}\sum\limits_{m = 0}^\infty  {\left( {\frac{{4m + 3}}{{2m + 2}}} \right)} P_{2m} \left( 0 \right)a_{\left( {2m + 1} \right)\omega }  + \frac{1}{{2\pi }}\sum\limits_{m = 1}^\infty  {\left( {4m + 1} \right)} P_{2m} \left( 0 \right)b_{\left( {2m} \right)\omega } , \nonumber \\ 
 T_\omega ^{eo}  &=& \frac{1}{{2\pi }}\sum\limits_{m = 1}^\infty  {\left( {4m + 1} \right)} P_{2m} \left( 0 \right)a_{\left( {2m} \right)\omega }  + \frac{1}{{2\pi }}\sum\limits_{m = 0}^\infty  {\left( {\frac{{4m + 3}}{{2m + 2}}} \right)P_{2m} \left( 0 \right)} b_{\left( {2m + 1} \right)\omega },
\end{eqnarray}
whose superscripts emphasize their dependencies on the Mie coefficients, i.e. $T_\omega ^{oe}$ depends on the odd $a_{n\omega }$ and the even $b_{n\omega }$ whereas $T_\omega ^{eo}$ depends on the even $a_{n\omega }$ and the odd $b_{n\omega }$. In turn the probability density of polariton coincidence detection in the first of Eqs.(\ref{Prob-Den}) becomes
\begin{equation} \label{Disp_Coin}
{\frac{{dP_{ss}^\sigma  \left( {\omega ^1 {\bf{N}}^1 \Lambda ^1 ,\omega ^2 {\bf{N}}^2 \Lambda ^2 } \right)}}{{d\omega ^1 do_{{\bf{N}}^1 } d\omega ^2 do_{{\bf{N}}^2 } }} = \left( {1 + \sigma } \right)^2 \frac{{\Delta o^2 }}{2}\left| {\begin{pmatrix}
   {\delta _{\Lambda ^1 1} } & {\delta _{\Lambda ^1 2} }  \\
\end{pmatrix}\begin{pmatrix}
   {  \phi _{11}^\sigma  \left( {\omega ^1 ,\omega ^2 } \right)\left( {T_\omega ^{oe} } \right)^2 } & {\phi _{12}^\sigma  \left( {\omega ^1 ,\omega ^2 } \right)T_\omega ^{oe} T_\omega ^{eo} }  \\
   {\phi _{21}^\sigma  \left( {\omega ^1 ,\omega ^2 } \right)T_\omega ^{oe} T_\omega ^{eo} } & {  \phi _{22}^\sigma  \left( {\omega ^1 ,\omega ^2 } \right)\left( {T_\omega ^{eo} } \right)^2 }  \\
\end{pmatrix}\begin{pmatrix}
   {\delta _{\Lambda ^2 1} }  \\
   {\delta _{\Lambda ^2 2} }  \\
\end{pmatrix}} \right|^2 },
\end{equation}
which evidently vanishes for $\sigma = -1$ since the here considered scattering geometry of Fig.3 belongs to the first class discussed in Sec.IV. Note that such probability density for each  polarization pair $\left( {\Lambda ^1 ,\Lambda ^2 } \right)$ is proportional to squared modulus of the corresponding entry in the inner matrix. 

By using the Legendre polynomials property $P'_n \left( { \pm 1} \right) = \left( { \pm 1} \right)^{n + 1} n\left( {n + 1} \right)/2$, the power scattering and absorption coefficients in the second and third of Eqs.(\ref{Sph_sSA}) for the considered scattering geometry become
\begin{eqnarray} \label{Sphere_geo_SA}
 S_{{\bf{m}}^j \lambda '}^{{\bf{m}}^i \lambda } \left( \omega  \right) &=& \delta _{ij} \delta _{\lambda \lambda '} L_\omega ^ +   + \left( {1 - \delta _{ij} } \right)\left( { - \delta _{\lambda 1} \delta _{\lambda '1}  + \delta _{\lambda 2} \delta _{\lambda '2} } \right)L_\omega ^ - , \nonumber  \\ 
 A_{{\bf{m}}^j \lambda '}^{{\bf{m}}^i \lambda } \left( \omega  \right) &=& \delta _{ij} \delta _{\lambda \lambda '} K_\omega ^ +   + \left( {1 - \delta _{ij} } \right)\left( { - \delta _{\lambda 1} \delta _{\lambda '1}  + \delta _{\lambda 2} \delta _{\lambda '2} } \right)K_\omega ^ -  ,
\end{eqnarray}
where 
\begin{eqnarray} \label{Disp_LpmKpm}
 L_\omega ^ \pm   &=& \frac{1}{{2\pi }}\sum\limits_{n = 1}^\infty  {\left( { \pm 1} \right)^{n + 1} \left( {2n + 1} \right)} \left( {\left| {a_{n\omega } } \right|^2  \pm \left| {b_{n\omega } } \right|^2 } \right), \nonumber \\ 
 K_\omega ^ \pm   &=& \frac{1}{{2\pi }}\sum\limits_{n = 1}^\infty  {\left( { \pm 1} \right)^{n + 1} \left( {2n + 1} \right)} \left\{ {\left[ {{\mathop{\rm Re}\nolimits} \left( {a_{n\omega } } \right) - \left| {a_{n\omega } } \right|^2 } \right] \pm \left[ {{\mathop{\rm Re}\nolimits} \left( {b_{n\omega } } \right) - \left| {b_{n\omega } } \right|^2 } \right]} \right\},
\end{eqnarray}
quantities whose physical meaning is directly provided by the discussion in Sec.IIIC. Specifically, from Eqs.(\ref{Sphere_geo_SA}), $L_\omega ^ +$ and $K_\omega ^ +$ are the diagonal terms of the power scattering and absorption coefficients and they represent the classical scattering and absorption cross-sections of a single plane wave, in agreement with Eqs.(\ref{Sphere_CscaSabs}), whereas $L_\omega ^ -$ and $K_\omega ^ -$ are the off-diagonal terms representing the classical interference of two plane waves in the considered scattering geometry. By exploiting the spectral overlap integral properties in Eqs.(\ref{OvIn_prop}), the polariton scattering probabilities of Eqs.(\ref{P2s_P1s_P0s_02}) become
\begin{eqnarray} \label{Disp_SurvProb}
 P_{2s}^\sigma   &=& \Delta o^2 \left\{ {\left( {L_\omega ^ +  } \right)^2  + \sigma \left[ {1 - 4\left( {I^\sigma  } \right)_{12}^{12} } \right]\left( {L_\omega ^ -  } \right)^2 } \right\}, \nonumber \\ 
 P_{1s}^\sigma   &=& 2\Delta o^2 \left\{ {L_\omega ^ +  K_\omega ^ +   + \sigma \left[ {1 - 4\left( {I^\sigma  } \right)_{12}^{12} } \right]L_\omega ^ -  K_\omega ^ -  } \right\}, \nonumber \\ 
 P_{0s}^\sigma   &=& \Delta o^2 \left\{ {\left( {K_\omega ^ +  } \right)^2  + \sigma \left[ {1 - 4\left( {I^\sigma  } \right)_{12}^{12} } \right]\left( {K_\omega ^ -  } \right)^2 } \right\},
\end{eqnarray}
whose structure, in view of the physical meaning of $L_\omega ^ \pm$ and $K_\omega ^ \pm$, unveils the role played by classical plane waves interference in the emergence of quantum interference effects. In fact, in each of the scattering probabilities of Eqs.(\ref{Disp_SurvProb}), the non-interference (first) contributions contain the correct number of single plane wave scattering and absorption cross-sections $L_\omega ^ +$ and $K_\omega ^ +$ whereas the interference (second) contributions, with the factor $\sigma$, contain the correct number of classical two plane waves interference terms $L_\omega ^ -$ and $K_\omega ^ -$. Note that the scattering probabilities of Eqs.(\ref{Disp_SurvProb}) explicitly depend on the ingoing wavepacket spectrum only through the overlap integral $\left( {I^\sigma  } \right)_{12}^{12}  = \int {d\omega ^1 d\omega ^2 } \left| {\phi _{12}^\sigma  \left( {\omega ^1 ,\omega ^2 } \right)} \right|^2$ appearing in their interference contributions.

\begin{figure}
\centering
\includegraphics[width = 1\linewidth]{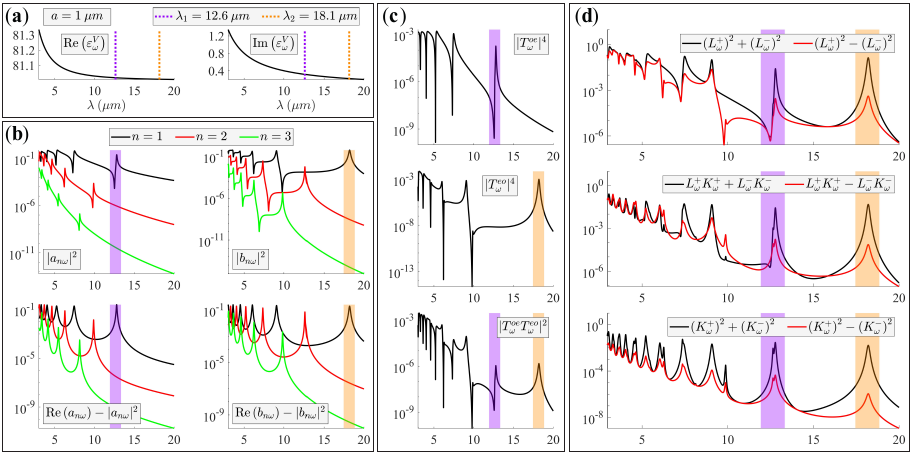}
\caption{Quantum optical scattering by a subwavelength-sized lossy sphere with high-index dielectric behavior. (a) Real and imaginary parts of the dielectric permitivity. (b) Square moduli $\left| {a_{n\omega } } \right|^2$, $\left| {b_{n\omega } } \right|^2$ and  combinations ${\mathop{\rm Re}\nolimits} \left( {a_{n\omega } } \right) - \left| {a_{n\omega } } \right|^2$, ${\mathop{\rm Re}\nolimits} \left( {b_{n\omega } } \right) - \left| {b_{n\omega } } \right|^2$ of the Mie coefficients. (c) Plots of the relevant quantities ruling the probability density of polariton coincidence detection of Eq.(\ref{Disp_Coin}). (d) Plots of the relevant quantities yielding the polariton scattering probabilities of Eqs.(\ref{Disp_SurvProb}).}
\label{Fig4}
\end{figure}

In order to explicitly discuss the spectral features of the polariton coincidence detection probability density in Eq.(\ref{Disp_Coin}) and of the polariton scattering probabilities of Eqs.(\ref{Disp_SurvProb}), we consider a sphere of radius $a = 1 \: \mu m$ whose Drude-Lorentz permittivity $\varepsilon _\omega ^V$, for wavelengths ranging from $3 \: \mu m$ to $20 \: \mu m$, has the real and imaginary parts plotted  in Fig.4(a). Accordingly, the subwavelength-sized sphere is weakly dispersive with marked dielectric behavior (high and positive permittivity real part) and moderate but non-negligible losses. The purple and orange dashed vertical lines locate the two specific wavelengths $\lambda_1 = 12.6 \: \mu m$  and $\lambda_2 = 18.1 \: \mu m$, respectively, which we exploit in the following spectral analysis. 

In Fig.4(b) we report, for $n=1,2,3$, the plots of the square moduli $\left| {a_{n\omega } } \right|^2$, $\left| {b_{n\omega } } \right|^2$ and of the combinations ${\mathop{\rm Re}\nolimits} \left( {a_{n\omega } } \right) - \left| {a_{n\omega } } \right|^2$, ${\mathop{\rm Re}\nolimits} \left( {b_{n\omega } } \right) - \left| {b_{n\omega } } \right|^2$ of the Mie coefficients, the latters assessing the overall impact of sphere losses on the scattering process, as discussed in Sec.IIID. The first striking feature of Fig.4(b) is that the quantities of order $n=3$ (green) are generally much smaller than those of order $n=1$ (black) and $n=2$ (red) so that all the higher order ones ($n \ge 4$, not plotted here) can be neglected in the present analysis. Moreover, Fig.4(b) shows that all the quantities globally decrease at larger wavelength (where the sphere is electromagnetically smaller) and, most importantly, that they exhibit various Mie resonance peaks whose spectral spacing increases at larger wavelengths. It is crucial to stress here that, due to the sphere high-index dielectric behavior, Mie resonance peaks of $\left| {a_{n\omega } } \right|^2$ and $\left| {b_{n\omega } } \right|^2$ exhibit characteristic Fano lineshapes \cite{Tzaro1} with different asymmetry degrees (tuned by the Fano parameter of the specific peak) and possibly displaying strong lateral dips where $\left| {a_{n\omega } } \right|^2$ and $\left| {b_{n\omega } } \right|^2$ drop down by several orders of magnitude (the dips reach zero only for ideally transparent spheres). Besides, Mie resonances are also accompanied by strong energy dissipation in the presence of matter losses, as evidenced by the corresponding peaks in ${\mathop{\rm Re}\nolimits} \left( {a_{n\omega } } \right) - \left| {a_{n\omega } } \right|^2$ and ${\mathop{\rm Re}\nolimits} \left( {b_{n\omega } } \right) - \left| {b_{n\omega } } \right|^2$ whose spectral profiles are practically symmetric since the Fano mechanism is inhibited in the absorption cross section. Note that the wavelengths $\lambda_1$ and $\lambda_2$ introduced in Fig.4(a) correspond to specific Mie resonances of $\left| {a_{1\omega } } \right|^2$ and $\left| {b_{1\omega } } \right|^2$ and we have highlighted the spectral regions surrounding such wavelengths by means of purple and orange shady rectangles, respectively. 

In Fig.4(c) we have plotted the combinations of the quantities $T_\omega ^{oe}$ and $T_\omega ^{eo}$ of Eqs.(\ref{Disp_ToeTeo}) appearing in the probability density of polariton coincidence detection of Eq.(\ref{Disp_Coin}) which, we repeat, vanishes for antisymmetric wavepackets ($\sigma = -1$) due to perfect destructive interference so that we here focus on symmetric wavepackets ($\sigma = +1$). For example, $\left| {T_\omega ^{oe} } \right|^4$ basically lends its spectral features to the probability density of coincide detection of the outgoing polaritons of directions ${\bf{N}}^1  =  - {\bf{u}}_y$, ${\bf{N}}^2  = {\bf{u}}_y$ and polarizations ${\bf{e}}_{{\bf{N}}^1 1}  = {\bf{e}}_{{\bf{N}}^2 1}  = {\bf{u}}_z$ (see Fig.3). The most important result shown by Fig.4(c) is that the Mie resonances yield marked spectral peaks in the probability density of coincidence detection of any pair of outgoing polaritons, which physically means that classical resonances are responsible for large quantum constructive interference. The spectral features of such peaks can easily be grasped after noting that, for the Mie coefficients in Fig.4(b), Eqs.(\ref{Disp_ToeTeo}) can be approximated as $T_\omega ^{oe}  \simeq \frac{3}{{4\pi }}P_0 \left( 0 \right)a_{1\omega }$ and $ T_\omega ^{eo}  \simeq \frac{3}{{4\pi }}P_0 \left( 0 \right)b_{1\omega }$ so that, for example, the peak at $\lambda_1$ (purple) of $\left| {T_\omega ^{oe} } \right|^4$ and at $\lambda_2$ (orange) of $\left| {T_\omega ^{eo} } \right|^4$ are produced by the corresponding Mie resonances of $a_{1\omega}$ and $b_{1\omega}$, respectively (and such two peaks evidently show up together in $\left| {T_\omega ^{oe} T_\omega ^{eo} } \right|^2$). Is is also worth noting that the possible asymmetric line-shapes of the Mie peaks in $\left| {a_{n\omega } } \right|^2$ and $\left| {b_{n\omega } } \right|^2$, physically due to the Fano mechanism, correspondingly yields significant dips in Fig.4(c), representing spectral regions where very large quantum destructive interference occurs,  as the one in $\left| {T_\omega ^{oe} } \right|^4$ which is close to the resonance wavelength $\lambda_1$ (purple) whose minimum is about six orders of magnitude smaller than the nearby maximum reached at resonance. It is also remarkable that the almost symmetric peak in $\left| {T_\omega ^{eo} } \right|^4$ at the wavelength $\lambda_2$ (orange) has a maximum which is about five orders of magnitude larger than its neighbor off-resonance background. 

In Fig.4(d) we analyze the polariton scattering probabilities of Eqs.(\ref{Disp_SurvProb}), for wavepackets with $\left( {I^\sigma  } \right)_{12}^{12} = 0$ (for the sake of simplicity), by plotting the quantities $\left( {L_\omega ^ +  } \right)^2  + \sigma \left( {L_\omega ^ -  } \right)^2$, $L_\omega ^ +  K_\omega ^ +   +\sigma L_\omega ^ -  K_\omega ^ - $ and $\left( {K_\omega ^ +  } \right)^2  +\sigma \left( {K_\omega ^ -  } \right)^2$, using black and red solid lines for  symmetric $\sigma = +1$ and antisymmetric $\sigma = -1$ wavepackets, respectively. Evidently, in close analogy with the polariton coincidence detection, also the polariton scattering probabilities exhibit marked peaks associated to the classical Mie resonances and such peaks can display lateral deep dips due to the Fano mechanism accompanying $\left| {a_{n\omega } } \right|^2$ and $\left| {b_{n\omega } } \right|^2$. In other words, classical Mie resonances have a large impact on quantum constructive and destructive interference effects accompanying the global polariton scattering processes. However, the results reported in Fig.4(d) are also interesting for two main additional reasons. First, the one and zero polariton scattering probabilities $P_{1s}^\sigma$ and $P_{0s}^\sigma$ are strictly vanishing for an ideally transparent sphere (and for any transparent object) so that their spectral features provide additional routes to perform quantum measurements which are not available when matter losses can be neglected. Second, Fig.4(d) makes it clear that the polariton scattering probabilities for symmetric (black lines) and antisymmetric (red lines) wavepackets are generally different, their discrepancy being very large at the resonance peaks of a single Mie coefficient, as the peaks within the orange shaded rectangles. Such large discrepancies occur since, close to resonances of this kind, the polariton survival probabilities of Eqs.(\ref{Disp_SurvProb}) for antisymmetric wavepackets $\sigma = -1$ are very small due to the physical fact that the interference terms $|L_\omega^-|$ and $|K_\omega^-|$ nearly equate the scattering and absorption cross-sections $|L_\omega^+|$ and $|K_\omega^+|$, respectively. As an example, the first of Eqs.(\ref{Disp_LpmKpm}) for the Mie coefficients in Fig.4(b) yields $ | {L_\omega ^ \pm  }  | \simeq \frac{3}{{2\pi }} | { | {a_{1\omega } }  |^2  \pm  | {b_{1\omega } }  |^2 }  |$ so that, close to the wavelength $\lambda _2$ where  $| {b_{1\omega } } |^2  \gg | {a_{1\omega } } |^2$ (i.e. only $b_{1\omega}$ resonates), we have that $\left| {L_\omega ^ +  } \right| \simeq \left| {L_\omega ^ -  } \right|$ and most importantly that 
\begin{equation}
\frac{{\left( {L_\omega ^ +  } \right)^2  - \left( {L_\omega ^ -  } \right)^2 }}{{\left( {L_\omega ^ +  } \right)^2  + \left( {L_\omega ^ -  } \right)^2 }} \simeq 2\frac{{\left| {a_{1\omega } } \right|^2 }}{{\left| {b_{1\omega } } \right|^2 }} \ll 1
\end{equation}
thus explaining the three order of magnitude discrepancy between the (red and black) peak values at the wavelength $\lambda _2$ (orange shaded rectangle) in the first subplot of Fig.4(d). We conclude that the quantum interference effects in the global polariton scattering processes produced by classical Mie resonances exhibit marked sensitivity to the symmetry of the wavepacket spectrum, a characteristic that could be exploited in an actual setup to efficiently detect polariton entanglement.

\section{Conclusion}
In conclusion we have exploited a recently introduced quantum optical scattering framework, based on the MLFN, to analyze the scattering of two directional $s$ polaritons by a mascroscopic lossy sphere which, as any other finite-size scatterer, can effectively be regarded as an optical device with an infinite number of 'input' and 'output' ports associated to the ingoing polariton directions. Classical matter losses physically yield the quantum mechanism of $s$ polariton absorption thus enabling the existence of three distinct processes where two, one or zero $s$ polaritons survive the scattering. If the ingoning $s$ polaritons are not collinear, the detections of such three processes display marked quantum interference effects which represent a far-reaching generalization of those supported by standard beam splitters since their efficient observation is facilitated by the much greater freedom of choosing the scattering geometry (i.e. the input and output ports). As an example, the equality of transmission and reflection rates is the only conditions enabling the beam splitter to exhibit the famous Hong-Ou-Mandel effect whereas we have here proved that different classes of scattering geometries allow the sphere to perform perfect destructive interference. Exploiting the MLFN ability to account for macroscopic matter dispersive response, we have discussed the impact of the sphere losses on the quantum interference effects arising in the detection of the scattered radiation. In particular we have shown that classical Mie resonances dramatically provide spectral bandwidths where very large constructive and destructive interference occurs. In analogy to the above discussed flexibility of the geometric setup provided by the sphere, also this result is a noteworthy generalization of the spectral abilities of lossy beam splitter since it usually does not exhibit resonances and it has only been shown to range from photon coalescence to anticoalescence. As a final remark, we stress that matter losses open up the possibility of performing measurements not allowed by ideally transparent objects as, for instance, the overall detection of one or zero scattered $s$ polaritons 
 in the two polaritons scattering considered in this paper. We have shown that the corresponding scattering probabilities are very sensitive to the spectral symmetry of the ingoing wavepacket when its central frequency matches an isolated Mie resonance. Since the discrepancy between the  scattering probabilities of symmetric and antisymmetric wavepackets can be of several order of magnitude, we believe that our results can suggest alternative strategies to detect polariton entanglement.

\appendix

\section{Classical electromagnetic scattering by a homogeneous lossy sphere}
In this appendix we evaluate the scattering coefficient 
$s_{{\bf{n}}\lambda }^{{\bf{N}}\Lambda } \left( \omega  \right)$ and the power scattering coefficients $S_{{\bf{n}}'\lambda '}^{{\bf{n}}\lambda }  \left( \omega  \right)$ and $A_{{\bf{n}}'\lambda '}^{{\bf{n}}\lambda } \left( \omega  \right)$ for a homogenous lossy sphere of radius $a$ and dielectric permittivity ${\varepsilon _\omega ^V }$; the overall dielectric permittivity is $\varepsilon _\omega  \left( {\bf{s}} \right) = {\rm H}\left( {s - a} \right) + {\rm H}\left( {a - s} \right)\varepsilon _\omega ^V$, where $s = |{\bf s}|$ and $\rm H$ is the Heasivisde step function. By resorting to the well-known Mie analytical treatment of plane wave scattering by a homogeneous sphere \cite{Jacks}, the modal dyadics ${\cal F}_\omega  \left( {\left. {\bf{s}} \right|{\bf{n}}} \right)$ satisfying the boundary value problem in the second row of Eqs.(\ref{GS}) is seen to be given by  
\begin{eqnarray} \label{Sph_MoDya}
&& {\cal F}_\omega  \left( {\left. {\bf{s}} \right|{\bf{n}}} \right) = {\rm H}\left( {s - a} \right) e^{ik_\omega  {\bf{s}} \cdot {\bf{n}}} {\cal I}_{\bf{n}} \nonumber \\ 
&&  + {\rm H}\left( {s - a} \right) 4\pi \sum\limits_{n = 1}^\infty  {\sum\limits_{m =  - n}^n {i^n \left\{ {-\frac{{a_{n\omega } }}{{i k_\omega  }}\nabla _{\bf{s}}  \times \left[ {h_n^{\left( 1 \right)} \left( {k_\omega  s} \right){\bf{X}}_{nm} \left( {\frac{{\bf{s}}}{s}} \right)} \right]{\bf{n}} \times {\bf{X}}_{nm}^* \left( {\bf{n}} \right) - b_{n\omega } h_n^{\left( 1 \right)} \left( {k_\omega  s} \right){\bf{X}}_{nm} \left( {\frac{{\bf{s}}}{s}} \right){\bf{X}}_{nm}^* \left( {\bf{n}} \right)} \right\}} }  \nonumber \\ 
&& + {\rm H}\left( {a - s} \right) 4 \pi \sum\limits_{n = 1}^\infty  {\sum\limits_{m =  - n}^n {i^n \left\{ {\frac{{d_{n\omega } }}{{ik_\omega ^V }}\nabla _{\bf{s}}  \times \left[ {j_n \left( {k_\omega ^V s} \right){\bf{X}}_{nm} \left( {\frac{{\bf{s}}}{s}} \right)} \right]{\bf{n}} \times {\bf{X}}_{nm}^* \left( {\bf{n}} \right) + c_{n\omega } j_n \left( {k_\omega ^V s} \right){\bf{X}}_{nm} \left( {\frac{{\bf{s}}}{s}} \right){\bf{X}}_{nm}^* \left( {\bf{n}} \right)} \right\}} } , 
\end{eqnarray}
where $k_\omega ^V  = k_\omega  \sqrt {\varepsilon _\omega ^V } $ (with ${{\mathop{\rm Im}\nolimits} ( {k_\omega ^V } ) \ge 0}$) is the wavenumber inside the sphere, $j_n \left( \rho  \right)$ and $h_n^{\left( 1 \right)} \left( \rho  \right)$ are the spherical Bessel and Hankel functions of the first kind,
\begin{equation}
{\bf{X}}_{nm} \left( {\bf{m}} \right) = \frac{1}{{\sqrt {n\left( {n + 1} \right)} }}\frac{1}{i}\left( { - {\bf{u}}_{\theta _{\bf{m}} } \frac{1}{{\sin \theta _{\bf{m}} }}\frac{\partial }{{\partial \varphi _{\bf{m}} }} + {\bf{u}}_{\varphi _{\bf{m}} } \frac{\partial }{{\partial \theta _{\bf{m}} }}} \right)Y_{nm} \left( {\theta _{\bf{m}} ,\varphi _{\bf{m}} } \right)
\end{equation}
is the vector spherical harmonic associated to the scalar spherical harmonic $Y_{nm}$ and 
\begin{eqnarray}
\begin{pmatrix}
   {a_{n\omega } }  \\
   {d_{n\omega } }  \\
\end{pmatrix} &=& \frac{{ \begin{pmatrix}
   {\displaystyle  j_n \left( {k_\omega ^V a} \right)\frac{\partial }{{\partial a}}\left[ {aj_n \left( {k_\omega  a} \right)} \right] - \frac{1}{\varepsilon _\omega ^V} j_n \left( {k_\omega  a} \right)\frac{\partial }{{\partial a}}\left[ {aj_n \left( {k_\omega ^V a} \right)} \right]}  \\
   { \displaystyle {j_n \left( {k_\omega ^V a} \right)\frac{\partial }{{\partial a}}\left[ {ah_n^{\left( 1 \right)} \left( {k_\omega ^V a} \right)} \right] - h_n^{\left( 1 \right)} \left( {k_\omega ^V a} \right)\frac{\partial }{{\partial a}}\left[ {aj_n \left( {k_\omega ^V a} \right)} \right]}     } , \nonumber \\
\end{pmatrix} }}{\displaystyle { j_n \left( {k_\omega ^V a} \right)\frac{\partial }{{\partial a}}\left[ {ah_n^{\left( 1 \right)} \left( {k_\omega  a} \right)} \right] - \frac{1}{\varepsilon _\omega ^V} h_n^{\left( 1 \right)} \left( {k_\omega  a} \right)\frac{\partial }{{\partial a}}\left[ {aj_n \left( {k_\omega ^V a} \right)} \right]}}, \\ 
\begin{pmatrix}
   {b_{n\omega } }  \\
   {c_{n\omega } }  \\
\end{pmatrix} &=& \frac{ \begin{pmatrix}
   {\displaystyle j_n \left( {k_\omega ^V a} \right)\frac{\partial }{{\partial a}}\left[ {aj_n \left( {k_\omega  a} \right)} \right] - j_n \left( {k_\omega  a} \right)\frac{\partial }{{\partial a}}\left[ {aj_n \left( {k_\omega ^V a} \right)} \right]}  \\
   {\displaystyle {j_n \left( {k_\omega  a} \right)\frac{\partial }{{\partial a}}\left[ {ah_n^{\left( 1 \right)} \left( {k_\omega  a} \right)} \right] - h_n^{\left( 1 \right)} \left( {k_\omega  a} \right)\frac{\partial }{{\partial a}}\left[ {aj_n \left( {k_\omega  a} \right)} \right]}}  \\
\end{pmatrix}}{\displaystyle {j_n \left( {k_\omega ^V a} \right)\frac{\partial }{{\partial a}}\left[ {ah_n^{\left( 1 \right)} \left( {k_\omega  a} \right)} \right] - h_n^{\left( 1 \right)} \left( {k_\omega  a} \right)\frac{\partial }{{\partial a}}\left[ {aj_n \left( {k_\omega ^V a} \right)} \right]}} 
\end{eqnarray}
are the standard Mie coefficients \cite{Bohre}. As a consistency check of Eq.(\ref{Sph_MoDya}), note that in the case where  there is no scattering sphere, i.e. $\varepsilon _\omega ^V  = 1$, the Mie coeffients become $a_{n\omega }  = b_{n\omega }  = 0$ and $d_{n\omega }  = c_{n\omega }  = 1$ so that the modal dyadics reduces to
\begin{equation}
 {\cal F}_\omega  \left( {\left. {\bf{s}} \right|{\bf{n}}} \right) = e^{ik_\omega  {\bf{s}} \cdot {\bf{n}}} {\cal I}_{\bf{n}} = 4 \pi \sum\limits_{n = 1}^\infty  {\sum\limits_{m =  - n}^n {i^n \left\{ {\frac{1}{{ik_\omega  }}\nabla _{\bf{s}}  \times \left[ {j_n \left( {k_\omega  s} \right){\bf{X}}_{nm} \left( {\frac{{\bf{s}}}{s}} \right)} \right]{\bf{n}} \times {\bf{X}}_{nm}^* \left( {\bf{n}} \right) + j_n \left( {k_\omega  s} \right){\bf{X}}_{nm} \left( {\frac{{\bf{s}}}{s}} \right){\bf{X}}_{nm}^* \left( {\bf{n}} \right)} \right\}} },
\end{equation}
correctly displaying a pure dyadic plane wave profile without scattered field (the second identity is the spherical expansion of the dyadic plane wave). It is worth pointing out that the Mie coefficients satisfy the remarkable relations
\begin{eqnarray}
 \frac{{{\mathop{\rm Re}\nolimits} \left( {a_{n\omega } } \right) - \left| {a_{n\omega } } \right|^2 }}{{\left| {d_{n\omega } } \right|^2 }} &=& \left( {k_\omega  a} \right){\mathop{\rm Im}\nolimits} \left\{ {\frac{{k_\omega ^V }}{{k_\omega ^{V*} }}j_n \left( {k_\omega ^V a} \right)\frac{\partial }{{\partial a}}\left[ {aj_n^* \left( {k_\omega ^V a} \right)} \right]} \right\}, \nonumber \\ 
 \frac{{{\mathop{\rm Re}\nolimits} \left( {b_{n\omega } } \right) - \left| {b_{n\omega } } \right|^2 }}{{\left| {c_{n\omega } } \right|^2 }} &=& \left( {k_\omega  a} \right){\mathop{\rm Im}\nolimits} \left\{ {j_n \left( {k_\omega ^V a} \right)\frac{\partial }{{\partial a}}\left[ {aj_n^* \left( {k_\omega ^V a} \right)} \right]} \right\},
\end{eqnarray}
whose relevance stems from the fact that their RHSs vanish when ${\mathop{\rm Im}\nolimits} \left( {k_\omega ^V } \right) = 0$ so that we conclude that the Mie coefficients of a transparent sphere are such that ${\mathop{\rm Re}\nolimits} \left( {a_{n\omega } } \right) = \left| {a_{n\omega } } \right|^2$ and ${\mathop{\rm Re}\nolimits} \left( {b_{n\omega } } \right) = \left| {b_{n\omega } } \right|^2$.

By exploiting the asymptotic behavior of the spherical Hankel function, $h_n^{\left( 1 \right)} \left( \rho  \right)\mathop  \approx \limits_{\rho  \to  + \infty } \left( { - i} \right)^{n + 1} \frac{{e^{i\rho } }}{\rho }$, the asymptotic expansion of the modal dyadic in Eq.(\ref{Sph_MoDya}) is straighforwardly seen to be
\begin{equation}
{\cal F}_\omega  \left( {\left. {s{\bf{N}}} \right|{\bf{n}}} \right)\mathop  \approx \limits_{s \to  + \infty } e^{i\left( {k_\omega  s} \right){\bf{N}} \cdot {\bf{n}}} {\cal I}_{\bf{n}}  + \frac{{e^{ik_\omega  s} }}{s}\frac{4 \pi i}{{k_\omega  }}\sum\limits_{n = 1}^\infty  {\sum\limits_{m =  - n}^n {\left[ {a_{n\omega } {\bf{N}} \times {\bf{X}}_{nm} \left( {\bf{N}} \right){\bf{n}} \times {\bf{X}}_{nm}^* \left( {\bf{n}} \right) + b_{n\omega } {\bf{X}}_{nm} \left( {\bf{N}} \right){\bf{X}}_{nm}^* \left( {\bf{n}} \right)} \right]} },
\end{equation}
whose comparison with the general asymptotic relation ${\cal F}_\omega  \left( {\left. {s{\bf{N}}} \right|{\bf{n}}} \right)\mathop  \approx \limits_{s \to  + \infty }  {e^{i\left( {k_\omega  s} \right){\bf{N}} \cdot {\bf{n}}} {\cal I}_{\bf{n}}  + \frac{{e^{ik_\omega  s} }}{s}{\cal S}_\omega  \left( {\left. {\bf{N}} \right|{\bf{n}}} \right)}$ (see the second of Eqs.(\ref{GS})) directly provides the scattering dyadic
\begin{equation}
{\cal S}_\omega  \left( {{\bf{N}}\left| {\bf{n}} \right.} \right) = \frac{{4\pi i}}{{k_\omega  }}\sum\limits_{n = 1}^\infty  {\sum\limits_{m =  - n}^n {\left[ {a_{n\omega } {\bf{N}} \times {\bf{X}}_{nm} \left( {\bf{N}} \right){\bf{n}} \times {\bf{X}}_{nm}^* \left( {\bf{n}} \right) + b_{n\omega } {\bf{X}}_{nm} \left( {\bf{N}} \right){\bf{X}}_{nm}^* \left( {\bf{n}} \right)} \right]} } 
\end{equation}
which, for notational convenience,  we write as
\begin{equation} \label{Sph_ScDya}
{\cal S}_\omega  \left( {{\bf{N}}\left| {\bf{n}} \right.} \right) = \frac{{4\pi i}}{{k_\omega  }}\sum\limits_{n = 1}^\infty  { \begin{pmatrix}
   a_{n\omega }  & b_{n\omega } 
\end{pmatrix} \sum\limits_{m =  - n}^n {\begin{pmatrix}
   {{\bf{N}} \times {\bf{X}}_{nm} \left( {\bf{N}} \right){\bf{n}} \times {\bf{X}}_{nm}^* \left( {\bf{n}} \right)}  \\
   {{\bf{X}}_{nm} \left( {\bf{N}} \right){\bf{X}}_{nm}^* \left( {\bf{n}} \right)}  \\
\end{pmatrix}} } .
\end{equation}
The orthogonality properties of the vector spherical harmonics
\begin{eqnarray}
 \int {do_{\bf{m}} \;} {\bf{X}}_{n'm'}^* \left( {\bf{m}} \right) \cdot {\bf{X}}_{nm} \left( {\bf{m}} \right) &=& \delta _{nn'} \delta _{mm'}, \nonumber \\ 
 \int {do_{\bf{m}} } {\bf{X}}_{n'm'}^* \left( {\bf{m}} \right) \cdot \left[ {{\bf{m}} \times {\bf{X}}_{nm} \left( {\bf{m}} \right)} \right] &=& 0, \nonumber \\ 
 \int {do_{\bf{m}} \;} \left[ {{\bf{m}} \times {\bf{X}}_{n'm'} \left( {\bf{m}} \right)} \right]^*  \cdot \left[ {{\bf{m}} \times {\bf{X}}_{nm} \left( {\bf{m}} \right)} \right] &=& \delta _{nn'} \delta _{mm'} 
\end{eqnarray}
enable to easily prove that
\begin{equation} \label{Sph_ScScDya}
\int {do_{\bf{N}} } {\cal S}_\omega ^{T*} \left( {\left. {\bf{N}} \right|{\bf{n}}} \right) \cdot {\cal S}_\omega  \left( {\left. {\bf{N}} \right|{\bf{n}}'} \right) = \left( {\frac{{4\pi }}{{k_\omega  }}} \right)^2 \sum\limits_{n = 1}^\infty  {\begin{pmatrix}
   {\left| {a_{n\omega } } \right|^2 } & {\left| {b_{n\omega } } \right|^2 }  \\
\end{pmatrix}\sum\limits_{m =  - n}^n { \begin{pmatrix}
   {{\bf{n}} \times {\bf{X}}_{nm} \left( {\bf{n}} \right){\bf{n}}' \times {\bf{X}}_{nm}^* \left( {{\bf{n}}'} \right)}  \\
   {{\bf{X}}_{nm} \left( {\bf{n}} \right){\bf{X}}_{nm}^* \left( {{\bf{n}}'} \right)}  \\
\end{pmatrix}} } .
\end{equation}
Besides, after inserting Eqs.(\ref{Sph_ScDya}) and (\ref{Sph_ScScDya}) into Eq.(\ref{Fund_Ene_Cons}) of the main text, we get
\begin{eqnarray} \label{Sph_AbAbDya}
\int {d^3 {\bf{R}}} {\cal A}_{\omega es}^{T*} \left( {{\bf{R}}\left| {\bf{n}} \right.} \right) \cdot {\cal A}_{\omega es} \left( {{\bf{R}}\left| {{\bf{n}}'} \right.} \right) = 4\sum\limits_{n = 1}^\infty  {\begin{pmatrix}
   {{\mathop{\rm Re}\nolimits} \left( {a_{n\omega } } \right) - \left| {a_{n\omega } } \right|^2 } & {{\mathop{\rm Re}\nolimits} \left( {b_{n\omega } } \right) - \left| {b_{n\omega } } \right|^2 }  \\
\end{pmatrix} \sum\limits_{m =  - n}^n {\begin{pmatrix}
   {{\bf{n}} \times {\bf{X}}_{nm} \left( {\bf{n}} \right){\bf{n}}' \times {\bf{X}}_{nm}^* \left( {{\bf{n}}'} \right)}  \\
   {{\bf{X}}_{nm} \left( {\bf{n}} \right){\bf{X}}_{nm}^* \left( {{\bf{n}}'} \right)}  \\
\end{pmatrix}} }. \nonumber \\
\end{eqnarray}
Now the addition theorem for spherical harmonics $\sum\limits_{m =  - n}^n {Y_{nm} \left( {\theta _{\bf{n}} ,\varphi _{\bf{n}} } \right)} Y_{nm}^* \left( {\theta _{{\bf{n}}'} ,\varphi _{{\bf{n}}'} } \right) = \frac{{2n + 1}}{{4\pi }}P_n \left( {{\bf{n}} \cdot {\bf{n}}'} \right)$,
where $P_n \left( \xi  \right)$ is the Legendre polynomial, enable to prove the dyadic addition formulae for the vector spherical harmonics
\begin{eqnarray} \label{VecSphHar_addit}
\sum\limits_{m =  - n}^n {\begin{pmatrix}
   {{\bf{m}} \times {\bf{X}}_{nm} \left( {\bf{m}} \right){\bf{n}} \times {\bf{X}}_{nm}^* \left( {\bf{n}} \right)}  \\
   {{\bf{X}}_{nm} \left( {\bf{m}} \right){\bf{X}}_{nm}^* \left( {\bf{n}} \right)}  \\
\end{pmatrix}}  =  \frac{1}{{4\pi }} \begin{pmatrix}
   {M_n^{11} \left( {{\bf{m}} \cdot {\bf{n}}} \right)} & {M_n^{12} \left( {{\bf{m}} \cdot {\bf{n}}} \right)}  \\
   {M_n^{21} \left( {{\bf{m}} \cdot {\bf{n}}} \right)} & {M_n^{22} \left( {{\bf{m}} \cdot {\bf{n}}} \right)}  \\
\end{pmatrix} \begin{pmatrix}
   {{\cal I}_{\bf{m}}  \cdot {\cal I}_{\bf{n}} }  \\
   {{\cal I}_{\bf{m}}  \cdot {\bf{nm}} \cdot {\cal I}_{\bf{n}} }  \\
\end{pmatrix},
\end{eqnarray}
where
\begin{equation}
\begin{pmatrix}
   {M_n^{11} \left( \xi  \right)} & {M_n^{12} \left( \xi  \right)}  \\
   {M_n^{21} \left( \xi  \right)} & {M_n^{22} \left( \xi  \right)}  \\
\end{pmatrix} = \frac{{2n + 1}}{{n\left( {n + 1} \right)}} \begin{pmatrix}
   {P'_n \left( \xi  \right)} & {P''_n \left( \xi  \right)}  \\
   { - \left( {1 - \xi ^2 } \right)P''_n \left( \xi  \right) + \xi P'_n \left( \xi  \right)} & { - \xi P''_n \left( \xi  \right) - P'_n \left( \xi  \right)}  \\
\end{pmatrix},
\end{equation}
which, inserted into Eqs.(\ref{Sph_ScDya}), (\ref{Sph_ScScDya}) and (\ref{Sph_AbAbDya}), yields
\begin{eqnarray}
 {\cal S}_\omega  \left( {{\bf{N}}\left| {\bf{n}} \right.} \right) &=& \frac{i}{{k_\omega  }}\sum\limits_{n = 1}^\infty  {   \begin{pmatrix}
   {a_{n\omega } } & {b_{n\omega } }  \\
\end{pmatrix} \begin{pmatrix}
   {M_n^{11} \left( {{\bf{N}} \cdot {\bf{n}}} \right)} & {M_n^{12} \left( {{\bf{N}} \cdot {\bf{n}}} \right)}  \\
   {M_n^{21} \left( {{\bf{N}} \cdot {\bf{n}}} \right)} & {M_n^{22} \left( {{\bf{N}} \cdot {\bf{n}}} \right)}  \\
\end{pmatrix}} \begin{pmatrix}
   {{\cal I}_{\bf{N}}  \cdot {\cal I}_{\bf{n}} }  \\
   {{\cal I}_{\bf{N}}  \cdot {\bf{nN}} \cdot {\cal I}_{\bf{n}} }  \\
\end{pmatrix}, \nonumber \\ 
 \int {do_{\bf{N}} } {\cal S}_\omega ^{T*} \left( {\left. {\bf{N}} \right|{\bf{n}}} \right) \cdot {\cal S}_\omega  \left( {\left. {\bf{N}} \right|{\bf{n}}'} \right) &=& \frac{{4\pi }}{{k_\omega ^2 }}\sum\limits_{n = 1}^\infty  { \begin{pmatrix}
   {\left| {a_{n\omega } } \right|^2 } & {\left| {b_{n\omega } } \right|^2 }  \\
\end{pmatrix} \begin{pmatrix}
   {M_n^{11} \left( {{\bf{n}} \cdot {\bf{n}}'} \right)} & {M_n^{12} \left( {{\bf{n}} \cdot {\bf{n}}'} \right)}  \\
   {M_n^{21} \left( {{\bf{n}} \cdot {\bf{n}}'} \right)} & {M_n^{22} \left( {{\bf{n}} \cdot {\bf{n}}'} \right)}  \\
\end{pmatrix} \begin{pmatrix}
   {{\cal I}_{\bf{n}}  \cdot {\cal I}_{{\bf{n}}'} }  \\
   {{\cal I}_{\bf{n}}  \cdot {\bf{n}}'{\bf{n}} \cdot {\cal I}_{{\bf{n}}'} }  \\
\end{pmatrix}} , \nonumber  \\ 
 \int {d^3 {\bf{R}}} {\cal A}_{\omega es}^{T*} \left( {{\bf{R}}\left| {\bf{n}} \right.} \right) \cdot {\cal A}_{\omega es} \left( {{\bf{R}}\left| {{\bf{n}}'} \right.} \right) &=& \frac{1}{\pi }\sum\limits_{n = 1}^\infty  {  \begin{pmatrix}
   {{\mathop{\rm Re}\nolimits} \left( {a_{n\omega } } \right) - \left| {a_{n\omega } } \right|^2 } & {{\mathop{\rm Re}\nolimits} \left( {b_{n\omega } } \right) - \left| {b_{n\omega } } \right|^2 }  \\
\end{pmatrix} \begin{pmatrix}
   {M_n^{11} \left( {{\bf{n}} \cdot {\bf{n}}'} \right)} & {M_n^{12} \left( {{\bf{n}} \cdot {\bf{n}}'} \right)}  \\
   {M_n^{21} \left( {{\bf{n}} \cdot {\bf{n}}'} \right)} & {M_n^{22} \left( {{\bf{n}} \cdot {\bf{n}}'} \right)}  \\
\end{pmatrix} \begin{pmatrix}
   {{\cal I}_{\bf{n}}  \cdot {\cal I}_{{\bf{n}}'} }  \\
   {{\cal I}_{\bf{n}}  \cdot {\bf{n}}'{\bf{n}} \cdot {\cal I}_{{\bf{n}}'} }  \\
\end{pmatrix}}. \nonumber  \\ 
\end{eqnarray}
Exploiting such dyadic expressions to evaluate  the scattering coefficient 
$s_{{\bf{n}}\lambda }^{{\bf{N}}\Lambda } \left( \omega  \right)$ in the first of Eqs.(\ref{sa}) and the power scattering coefficients $S_{{\bf{n}}'\lambda '}^{{\bf{n}}\lambda }  \left( \omega  \right)$ and $A_{{\bf{n}}'\lambda '}^{{\bf{n}}\lambda } \left( \omega  \right)$ of Eqs.(\ref{SA}), bearing in mind that ${\cal I}_{\bf{m}}  \cdot {\bf{e}}_{{\bf{m}}\lambda }  = {\bf{e}}_{{\bf{m}}\lambda }$, we get
\begin{eqnarray}
 s_{{\bf{n}}\lambda }^{{\bf{N}}\Lambda } \left( \omega  \right) &=& \frac{1}{\pi }\sum\limits_{n = 1}^\infty  {\begin{pmatrix}
   { - \frac{1}{2}a_{n\omega } } & { - \frac{1}{2}b_{n\omega } }  \\
\end{pmatrix}\begin{pmatrix}
   {M_n^{11} \left( {{\bf{N}} \cdot {\bf{n}}} \right)} & {M_n^{12} \left( {{\bf{N}} \cdot {\bf{n}}} \right)}  \\
   {M_n^{21} \left( {{\bf{N}} \cdot {\bf{n}}} \right)} & {M_n^{22} \left( {{\bf{N}} \cdot {\bf{n}}} \right)}  \\
\end{pmatrix}} \begin{pmatrix}
   {{\bf{e}}_{{\bf{N}}\Lambda }  \cdot {\bf{e}}_{{\bf{n}}\lambda } }  \\
   { {{\bf{e}}_{{\bf{N}}\Lambda }  \cdot {\bf{n}}}  {      {\bf{N}} \cdot {\bf{e}}_{{\bf{n}}\lambda }} }  \\
\end{pmatrix}, \nonumber \\ 
 S_{{\bf{n}}'\lambda '}^{{\bf{n}}\lambda } \left( \omega  \right) &=& \frac{1}{\pi }\sum\limits_{n = 1}^\infty  {\begin{pmatrix}
   {\left| {a_{n\omega } } \right|^2 } & {\left| {b_{n\omega } } \right|^2 }  \\
\end{pmatrix}\begin{pmatrix}
   {M_n^{11} \left( {{\bf{n}} \cdot {\bf{n}}'} \right)} & {M_n^{12} \left( {{\bf{n}} \cdot {\bf{n}}'} \right)}  \\
   {M_n^{21} \left( {{\bf{n}} \cdot {\bf{n}}'} \right)} & {M_n^{22} \left( {{\bf{n}} \cdot {\bf{n}}'} \right)}  \\
\end{pmatrix}\begin{pmatrix}
   {{\bf{e}}_{{\bf{n}}\lambda }  \cdot {\bf{e}}_{{\bf{n}}'\lambda '} }  \\
   { {{\bf{e}}_{{\bf{n}}\lambda }  \cdot {\bf{n}}'}   {{\bf{n}}}\cdot  {\bf{e}}_{{\bf{n}}'\lambda '}}  \\
\end{pmatrix}}, \nonumber  \\ 
 A_{{\bf{n}}'\lambda '}^{{\bf{n}}\lambda } \left( \omega  \right) &=& \frac{1}{\pi }\sum\limits_{n = 1}^\infty  {\begin{pmatrix}
   {{\mathop{\rm Re}\nolimits} \left( {a_{n\omega } } \right) - \left| {a_{n\omega } } \right|^2 } & {{\mathop{\rm Re}\nolimits} \left( {b_{n\omega } } \right) - \left| {b_{n\omega } } \right|^2 }  \\
\end{pmatrix}\begin{pmatrix}
   {M_n^{11} \left( {{\bf{n}} \cdot {\bf{n}}'} \right)} & {M_n^{12} \left( {{\bf{n}} \cdot {\bf{n}}'} \right)}  \\
   {M_n^{21} \left( {{\bf{n}} \cdot {\bf{n}}'} \right)} & {M_n^{22} \left( {{\bf{n}} \cdot {\bf{n}}'} \right)}  \\
\end{pmatrix}\begin{pmatrix}
   {{\bf{e}}_{{\bf{n}}\lambda }  \cdot {\bf{e}}_{{\bf{n}}'\lambda '} }  \\
    { {{\bf{e}}_{{\bf{n}}\lambda }  \cdot {\bf{n}}'}   {{\bf{n}}}\cdot  {\bf{e}}_{{\bf{n}}'\lambda '}} \\
\end{pmatrix}}, \nonumber  \\ 
\end{eqnarray}
which we rewrite in the more concise and expressive form
\begin{eqnarray}
 s_{{\bf{n}}\lambda }^{{\bf{N}}\Lambda } \left( \omega  \right) &=& {\rm J}_{{\bf{N}} \cdot {\bf{n}}} \left\{ {\begin{pmatrix}
   { - \frac{1}{2}a_{n\omega } } & { - \frac{1}{2}b_{n\omega } }  \\
\end{pmatrix}} \right\}\begin{pmatrix}
   {{\bf{e}}_{{\bf{N}}\Lambda }  \cdot {\bf{e}}_{{\bf{n}}\lambda } }  \\
   {{\bf{e}}_{{\bf{N}}\Lambda }  \cdot {\bf{nN}} \cdot {\bf{e}}_{{\bf{n}}\lambda } }  \\
\end{pmatrix}, \nonumber \\ 
 S_{{\bf{n}}'\lambda '}^{{\bf{n}}\lambda } \left( \omega  \right) &=& {\rm J}_{{\bf{n}} \cdot {\bf{n}}'} \left\{ {\begin{pmatrix}
   {\left| {a_{n\omega } } \right|^2 } & {\left| {b_{n\omega } } \right|^2 }  \\
\end{pmatrix}} \right\}\begin{pmatrix}
   {{\bf{e}}_{{\bf{n}}\lambda }  \cdot {\bf{e}}_{{\bf{n}}'\lambda '} }  \\
   {{\bf{e}}_{{\bf{n}}\lambda }  \cdot {\bf{n}}'{\bf{n}} \cdot {\bf{e}}_{{\bf{n}}'\lambda '} }  \\
\end{pmatrix}, \nonumber \\ 
 A_{{\bf{n}}'\lambda '}^{{\bf{n}}\lambda } \left( \omega  \right) &=& {\rm J}_{{\bf{n}} \cdot {\bf{n}}'} \left\{ {\begin{pmatrix}
   {{\mathop{\rm Re}\nolimits} \left( {a_{n\omega } } \right) - \left| {a_{n\omega } } \right|^2 } & {{\mathop{\rm Re}\nolimits} \left( {b_{n\omega } } \right) - \left| {b_{n\omega } } \right|^2 }  \\
\end{pmatrix}} \right\}\begin{pmatrix}
   {{\bf{e}}_{{\bf{n}}\lambda }  \cdot {\bf{e}}_{{\bf{n}}'\lambda '} }  \\
   {{\bf{e}}_{{\bf{n}}\lambda }  \cdot {\bf{n}}'{\bf{n}} \cdot {\bf{e}}_{{\bf{n}}'\lambda '} }  \\
\end{pmatrix},
\end{eqnarray}
where
\begin{equation}
{\rm J}_\xi  \left\{ {\begin{pmatrix}
   {p_n } & {q_m }  \\
\end{pmatrix}} \right\} = \frac{1}{\pi }\sum\limits_{n = 1}^\infty  {\frac{{2n + 1}}{{n\left( {n + 1} \right)}}} \begin{pmatrix}
   {p_n } & {q_m }  \\
\end{pmatrix}\begin{pmatrix}
   {P'_n \left( \xi  \right)} & {P''_n \left( \xi  \right)}  \\
   { - \left( {1 - \xi ^2 } \right)P''_n \left( \xi  \right) + \xi P'_n \left( \xi  \right)} & { - \xi P''_n \left( \xi  \right) - P'_n \left( \xi  \right)}  \\
\end{pmatrix}.
\end{equation}


\begin{thebibliography}{10}
\bibitem{Symmet} Examples of wavepackets with definite spectral symmetry which are particularly relavant in quantum optical applications are provided by the Bell states (without polarization entanglement) whose spectra are $ \varphi _{\lambda ^1 \lambda ^2 }^ \pm  \left( {\omega ^1 ,\omega ^2 } \right) = \frac{1}{{\sqrt 2 }}\left[ {\xi \left( {\omega ^1 } \right)\xi \left( {\omega ^2 } \right) \pm \zeta \left( {\omega ^1 } \right)\zeta \left( {\omega ^2 } \right)} \right]\delta _{\lambda ^1 1} \delta _{\lambda ^2 1}$ and $\psi _{\lambda ^1 \lambda ^2 }^ \pm  \left( {\omega ^1 ,\omega ^2 } \right) = \frac{1}{{\sqrt 2 }}\left[ {\xi \left( {\omega ^1 } \right)\zeta \left( {\omega ^2 } \right) \pm \zeta \left( {\omega ^1 } \right)\xi \left( {\omega ^2 } \right)} \right]\delta _{\lambda ^1 1} \delta _{\lambda ^2 1}$ with $\int {d\omega } \left| {\xi \left( \omega  \right)} \right|^2  = \int {d\omega } \left| {\zeta \left( \omega  \right)} \right|^2  = 1$ and $\int {d\omega } \xi ^* \left( \omega  \right)\zeta \left( \omega  \right) = 0$; evidently $\varphi ^ +$, $\varphi ^ -  $ and $\psi ^ + $ are symmetric whereas $\psi ^ - $ is antisymmetric.
\end{thebibliography}
\end{document}